\documentclass[fleqn,usenatbib,useAMS]{mnras}

\usepackage{xcolor}
\usepackage{comment}
\usepackage{amsmath}
\definecolor{color}{RGB}{25,25,112}
\definecolor{negro}{RGB}{0,0,0}
\definecolor{colorurl}{RGB}{25,25,112}
 \usepackage{orcidlink}
\usepackage{amssymb}
\usepackage{longtable}
\usepackage[T1]{fontenc}
\DeclareRobustCommand{\VAN}[3]{#2}
\let\VANthebibliography\thebibliography
\def\thebibliography{\DeclareRobustCommand{\VAN}[3]{##3}\VANthebibliography}
\usepackage{graphicx}	% Including figure files
\usepackage{supertabular,booktabs,capt-of}

\title[Analysis from TAROT, COATLI, and RATIR Observations]{Understanding the Nature of the Optical Emission in Gamma-Ray Bursts: Analysis from TAROT, COATLI, and RATIR Observations}

\author[Becerra et al.]{R.~L.~Becerra\,\orcidlink{0000-0002-0216-3415},$^{1}$\thanks{E-mail: rosa.becerra@correo.nucleares.unam.mx (RLB)},
A.~Klotz\,\orcidlink{0000-0002-2652-0069},$^{2}$
J.~L.~Atteia\,\orcidlink{0000-0001-7346-5114},$^{2}$
D.~Guetta\,\orcidlink{0000-0002-7349-1109},$^{3}$
A.~M.~Watson\,\orcidlink{0000-0002-2008-6927}$^{4}$
F.~De~Colle\,\orcidlink{0000-0002-3137-4633},$^{1}$
\newauthor
C.~Angulo-Valdez\,\orcidlink{0009-0002-6667-3294},$^{4}$
N.~R.~Butler\,\orcidlink{0000-0002-9110-6673},$^{5}$
S.~Dichiara\,\orcidlink{0000-0001-6849-1270},$^{6}$
N.~Fraija\,\orcidlink{0000-0002-0173-6453},$^{4}$
K.~Garcia-Cifuentes\,\orcidlink{0009-0001-2607-6359},$^{1}$\newauthor
A.~S.~Kutyrev\,\orcidlink{0000-0002-2715-8460},$^{7,8}$
W.~H.~Lee\,\orcidlink{0000-0002-2467-5673},$^{4}$
M.~Pereyra\,\orcidlink{0000-0001-6148-6532},$^{9}$
and
E.~Troja\,\orcidlink{0000-0002-1869-7817}$^{10}$\\
$^{1}$ Instituto de Ciencias Nucleares, Universidad Nacional Aut\'onoma de M\'exico, Apartado Postal 70-264, 04510 M\'exico, CDMX, Mexico\\
$^{2}$ IRAP, Universit\'e de Toulouse, CNRS, CNES, UPS, (Toulouse), France\\
$^{3}$Department of Physics, Ariel University, Ariel, Israel\\
$^{4}$ Instituto de Astronom{\'\i}a, Universidad Nacional Aut\'onoma de M\'exico, Apartado Postal 70-264, 04510 M\'exico, CDMX, Mexico\\
$^{5}$ School of Earth and Space Exploration, Arizona State University, Tempe, AZ 85287, USA\\
$^{6}$ Department of Astronomy and Astrophysics, The Pennsylvania State University, 525 Davey Lab, University Park, PA 16802, USA \\
$^{7}$ Department of Astronomy, University of Maryland, College Park, MD 20742-4111, USA\\
$^{8}$ Astrophysics Science Division, NASA Goddard Space Flight Center, 8800 Greenbelt Road, Greenbelt, MD 20771, USA\\
$^{9}$ CONACYT, Instituto de Astronom{\'\i}a, Universidad Nacional Aut\'onoma de M\'exico, 22860 Ensenada, BC, Mexico\\
$^{10}$ Department of Physics, University of Rome - Tor Vergata, via della Ricerca Scientifica 1, 00100 Rome, IT
}

\date{Accepted XXX. Received YYY; in original form ZZZ}

\pubyear{2023}

\begin{document}
\label{firstpage}
\pagerange{\pageref{firstpage}--\pageref{lastpage}}
\maketitle

\begin{abstract}
We collected the optical light curve data of 227 gamma-ray bursts (GRBs) observed with the TAROT, COATLI, and RATIR telescopes. These consist of 133 detections and 94 upper limits. We constructed average light curves in the observer and rest frames in both X-rays (from {\itshape Swift}/XRT) and in the optical. Our analysis focused on investigating the observational and intrinsic properties of GRBs. Specifically, we examined observational properties, such as the optical brightness function of the GRBs at $T=1000$ seconds after the trigger, as well as the temporal slope of the afterglow.
We also estimated the redshift distribution for the GRBs within our sample. Of the 227 GRBs analysed, we found that 116 had a measured redshift. Based on these data, we calculated a local rate of $\rho_0=0.2$ Gpc$^{-3}$ yr$^{-1}$ for these events with $z<1$.
To explore the intrinsic properties of GRBs, we examined the average X-ray and optical light curves in the rest frame. 
We use the {\scshape afterglowpy} library to generate synthetic curves to constrain the parameters typical of the bright GRB jet, such as energy (${\langle} {E_{0}}{\rangle}\sim 10^{53.6}$~erg), opening angle (${\langle}\theta_\mathrm{core}{\rangle}\sim 0.2$~rad), and density (${\langle}n_\mathrm{0}{\rangle}\sim10^{-2.1}$ cm$^{-3}$). Furthermore, we analyse microphysical parameters, including the fraction of thermal energy in accelerated electrons (${\langle}\epsilon_e{\rangle}\sim 10^{-1.37}$) and in the magnetic field (${\langle}\epsilon_B{\rangle}\sim10^{-2.26}$), and the power-law index of the population of non-thermal electrons (${\langle}p{\rangle}\sim 2.2$).

\end{abstract}

\begin{keywords}

(stars:) gamma-ray burst: individual: GRB 180706A,
(stars:) gamma-ray burst: individual: GRB 180812A,
(transients:) gamma-ray bursts
\end{keywords}

\section{Introduction}
\label{sec:introduction}

Gamma-ray bursts (GRBs) are the brightest electromagnetic events observed in the universe \citep{OConnor2023,Atteia2017}. These bursts can be classified into two distinct populations: long GRBs (LGRBs) and short GRBs (SGRBs) \citep{Kouveliotou1993}. This classification is based primarily on the duration $T_{90}$, which represents the time interval that encompasses 90\% of the total observed counts in gamma rays. LGRBs (with $T_{90}\gtrsim2$ s) are associated with the collapse of massive stars \citep[e.g.,][]{Woosley1993,MacFadyen1999} whereas SGRBs (with $T_{90}\lesssim 2$ s) are believed to result from the coalescence of two compact objects \citep[e.g.,][]{Paczynski1986,Paczynski1991,Lee2007,Abbott2017}.

The fireball model \citep{Sari1998} is the standard theory used to explain most of the features observed in the GRB light curves. This model describes both the prompt emission and the afterglow phase. During the prompt phase, the rapid variability is produced by internal shocks in the jet \citep{Rees1994,Kobayashi1997}, by photospheric emission \citep{Thompson1994,Eichler2000}, or by magnetic reconnection \citep{Zhang2011,Metzger2011}. As the jet interacts with the circumburst medium, it decelerates, generating the afterglow phase. This phase is characterised by synchrotron emission, where relativistic electrons are continuously accelerated by an ongoing shock that interacts with the surrounding medium \citep{Granot2002}. This blast wave creates a forward shock that propagates through the surrounding medium and a reverse shock that propagates through the shell \citep{Sari1998}.

The emission of the afterglow can be described by power-law segments of the flux $F \propto t^{-\alpha}\nu^{-\beta}$, where $T$, $\nu$, $\alpha$, and $\beta$ represent time, frequency, temporal index, and spectral index \citep{Granot2002,Sari1998}.

With the launch of the {\itshape Swift} mission in 2004, our understanding of afterglows has greatly expanded \citep{Gehrels2004}. This is largely due to the monitoring provided by the X-Ray Telescope (XRT), which has provided an abundance of data from the very beginning of the afterglow. Unlike lower frequencies, the large amount of X-ray data has enabled both statistical studies and a in-depth investigations of the light curves of GRBs at these energies \citep{Burrows2005}.

The observational phases of X-ray afterglows can generally be identified as follows: a steep decay phase related to the end of the prompt emission phase; a shallow decay phase (or plateau) commonly interpreted as a signature of late-time energy injection; a normal decay phase; and a jet-break steepening given by a geometric effect related with the observer seeing the edge of the jet at late times. Additionally, flares are sometimes observed \citep[see, e.g.][]{Zhang2006,Nousek2006,Evans2009}. These flares are believed to be produced by the central engine in a manner analogous to the prompt emission \citep{Burrows2005}.

However, observing the GRB early optical emission by ground telescopes is not an easy task. This is because gamma-ray bursts (in their prompt emission phase) are short-lived, lasting only a few seconds to a few minutes, and most ground-based telescopes simply cannot respond quickly enough to capture their first seconds. Moreover, the telemetry delays in receiving alerts from the General Coordinates Network/Transient Astronomy Network (GCN/TAN)\footnote{\url{https://gcn.gsfc.nasa.gov/about.html}},
result in the loss of early information from GRBs at these frequencies.

Therefore, the sample size of the early optical phases of the GRB remains considerably smaller compared to the later afterglows \citep[see e.g.][]{Wang2013,Li2012,Kann2010}. 
Despite this limitation, these samples provide valuable insight into the diverse range of apparent brightness values observed during the early phases of GRBs.
In some cases, extremely faint nearby GRBs have been observed \citep[see e.g.][]{Liang2007} while, in other cases, the observed brightness is much higher than expected \citep[see e.g.][]{Perley2014,OConnor2023,Becerra2023}, reaching apparent magnitudes of approximately $r = 14$--16 \citep[see e.g.][]{Pereyra2022,Klotz2009}. This suggests that the observed brightness is not solely dependent on the distance.

Moreover, not all GRBs have an optical counterpart, and the reason for this darkness has been widely discussed \citep[see e.g.][]{Greiner2011}. Possible explanations have been proposed by \cite{Lang2010}: i) the intrinsically low luminosity of the GRB; ii) the environment surrounding the GRB; and iii) the high absorption of intergalactic medium due to large distances ($z>6$).

Small aperture, fast slewing telescopes are usually used for observations of optical counterparts of GRBs \citep[see e.g.][]{Klotz2006,Becerra2019a}, whereas larger aperture telescopes are necessary for detecting the afterglow when it has faded. In this paper, following this pattern, we use early optical data from the COATLI and TAROT telescopes and late data from RATIR \citep[see, e.g.][]{Becerra2019b}.
We aim to describe the observational properties of a sample of early optical afterglows followed by COATLI, TAROT, and RATIR. We describe each of them in \S~\ref{sec:telescopes} and the complete sample in \S~\ref{sec:sample}. We present the phenomenological results in \S~\ref{sec:obsresults}: the brightness function at $T+1000$~s and the subsequent analysis at the early phases of the afterglow. In this section, we also compare the X-ray and optical flux to classify the events of our sample as dark or not. Additionally, we analyse the redshift distribution of our sample in \S~\ref{sec:phyresults}.
Finally, we discuss and summarise our conclusions in \S~\ref{sec:discussion}. 

\section{Telescopes}
\label{sec:telescopes}

\subsection{TAROT}
\label{sec:tarot}

The T\'elescopes \`a Action Rapide pour les Objets Transitoires (TAROT)\footnote{\url{http://tarot.obs-hp.fr/}} are an automated telescope network located at different sites around the world with automated observations (without human interaction). TAROT was first designed as a robotic observatory in 1995. The Calern (TCA) observatory in France hosted the first TAROT, and its first light was in 1998. TAROT subsequently expanded to two other sites: La Silla, Chile (TCH) and La Reunión (TRE). TCA and TCH have a FoV 1.8$\times$ 1.8 square degrees, whereas the FoV of TRE is 4.2$\times$ 4.2 square degrees. TCA and TCH have a limiting $R$ magnitude of $R=18.2$ at 5$\sigma$ for unfiltered exposures of one minute, whereas the limiting magnitude of TRE is magnitudes of $R = 17$.
The goal of this consortium is the very early observations of GRB optical counterparts \citep{Klotz2006}. TAROT is connected to the GCN/TAN alert system and its first exposure is trailed with a duration of 60~seconds to allow continuous monitoring of the light curve. For these exposure, the tracking the hour-angle motor was adapted to give a drift of 0.30~pixels/s. As a consequence, stars trail with a length of about 18~pixels on the image and the flux is recorded continuously without dead time
\citep{Klotz2006}. Subsequent exposures are typically conventional and do have a small amount of dead time.

%GRB 071109 GCN 7052, trigger por INTEGRAL 4761
%GRB 071117 GCN 7108 datos no muy confiables 
%GRB 081003 GCN 8326, trigger por INTEGRAL 5361
%GRB 091202 GCN 10236, trigger por INTEGRAL 5957
%GRB 110206A GCN 11656, trigger por INTEGRAL 6608
%GRB 120202A GCN 12910, trigger por INTEGRAL 6461
%GRB 131218A GCN 15602, trigger por INTEGRAL 6409
%GRB 160629A GCN 19622, trigger por INTEGRAL 7498
%GRB 160629A GCN 19622, trigger por INTEGRAL 7498
%GRB 161023A GCN 20110, trigger por INTEGRAL 7613
%GRB 161214A GCN 20254, trigger por INTEGRAL 7644
% GRB 170409A detected by Fermi LAT and GBM (trigger 513398525/170409112) GCN 21008
%GRB 170423A detected by Fermi GBM (trigger 514660513 / 170423719) GCN 21076

\subsection{COATLI}
\label{sec:coatli}

COATLI\footnote{\url{http://coatli.astroscu.unam.mx/}} is an ASTELCO 50-cm Richey-Crétien telescope on a fast ASTELCO NTM-500 German equatorial mount at the Observatorio Astron\'omico Nacional on Sierra de San Pedro M\'artir in Baja California, Mexico \citep{Watson2016,Cuevas2016}.

COATLI is connected to the GCN/TAN alert system and observes GRBs autonomously. The reduction pipeline \citep{Becerra2019a} performs bias subtraction, dark subtraction, flat-field correction, and cosmic ray cleaning with the \emph{cosmicrays} task in {\scshape iraf} \citep{Tody1986}. We perform astrometric calibration of our images using the \url{astrometry.net} software \citep{Lang2010}. We calibrate against USNO-B1 \citep{Monet2003} and Pan-STARRS DR1 \citep{Magnier2020}.

The COATLI $w$ photometry is well described by $(w - r') = 0.353(g' - r') - 0.243{(g' - r')}^2$ where the apostrophes in $g'$ and $r'$ refer to SDSS magnitudes \citep{Becerra2019a}.  

\subsection{RATIR}
\label{sec:ratir}

The Reonization and Transients Infrared/Optical Project (RATIR)\footnote{\url{http://ratir.astroscu.unam.mx/}} was a four-channel simultaneous optical and near-infrared imager mounted on the 1.5~meters Harold L. Johnson Telescope at the Observatorio Astron\'omico Nacional in Sierra San Pedro M\'artir in Baja California, Mexico. It was operated from 2012 April to 2022 June. RATIR responded autonomously to GRB triggers from the {\itshape Swift}/BAT instrument and obtained simultaneous photometry in $riZJ$ or $riYH$ \citep{Butler2012,Watson2012,Littlejohns2015,Becerra2017}. 

The reduction pipeline \citep{Littlejohns2015} performs bias subtraction and flat-field correction, followed by astrometric calibration using the \url{astrometry.net} software \citep{Lang2010}, iterative sky subtraction, co-addition using {\scshape swarp} \citep{Bertin2010}, and source detection using {\scshape sextractor} \citep{Bertin1996}. We calibrate against SDSS DR9 \citep{Ahn2012} and 2MASS \citep{Skrutskie2006}.
 
\section{Data sample}
\label{sec:sample}

We used all available optical photometric data from TAROT, RATIR, and COATLI. TAROT and COATLI use clear filters, whereas the RATIR \emph{ griZYJH} photometry system. For typical GRBs, the AB magnitudes in the TAROT and COATLI clear filters ($w$-filters) are approximately the same as those in the RATIR $r$-filter \citep[see e.g.][]{Becerra2019b,Becerra2019c,Becerra2021}.

\begin{figure}
%\centering
 \includegraphics[width=0.45\textwidth]{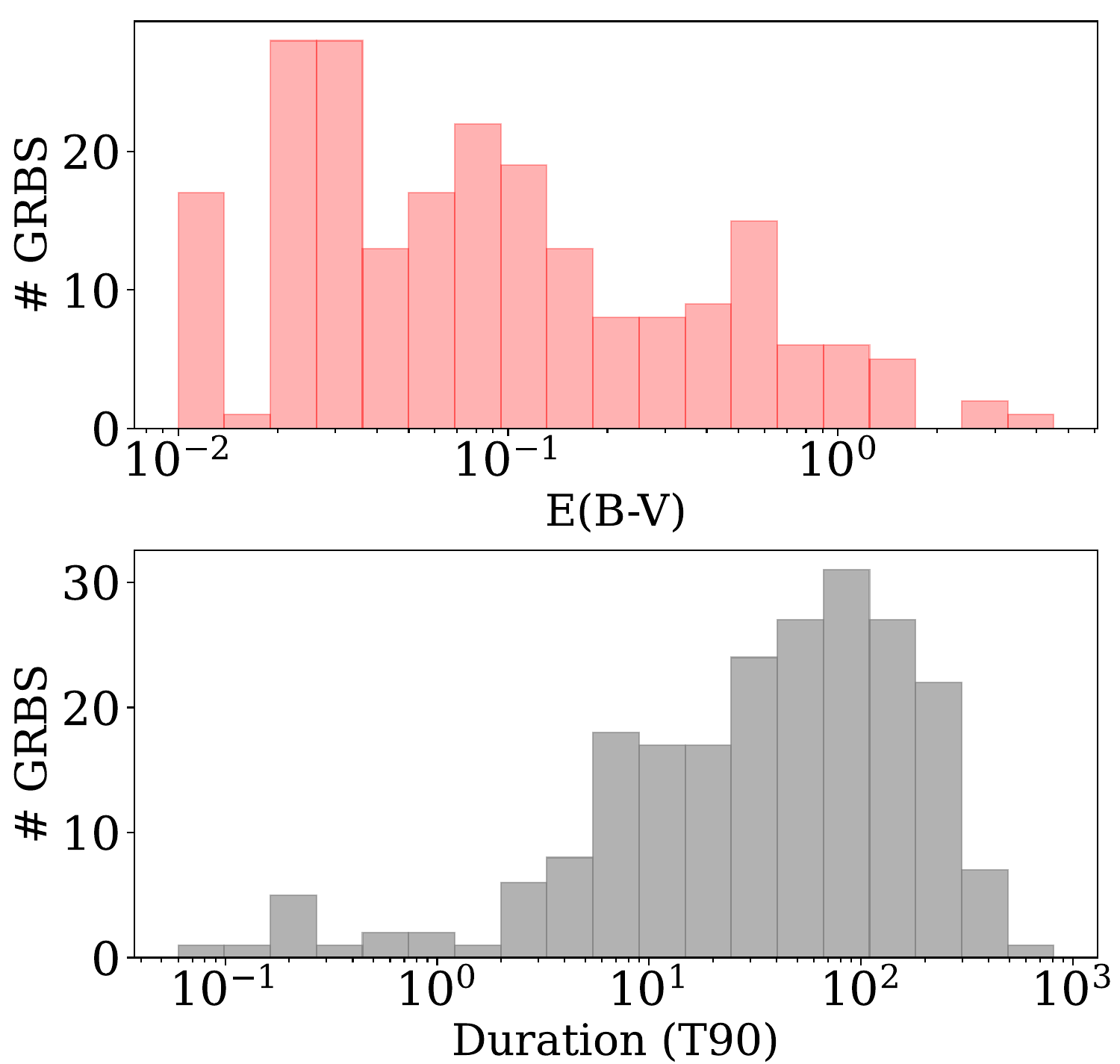}
 \caption{Histograms of reddening values $E(B-V)$ (top panel), and duration $T_{90}$ (bottom panel).}
 \label{fig:hist2}
\end{figure}

The complete information for each event is described in Table~3. The first column shows the name of the GRB, the second the {\itshape Swift}/BAT trigger, and the third, fourth, and fifth columns list the redshift \emph{z}, the extinction value $E(B-V)$, and the duration $T_{90}$, respectively. The information on the redshift $z$, reddening $E(B-V)$, and duration $T_{90}$ was retrieved from the GCN/TAN circulars and the {\itshape Swift}/BAT repository\footnote{\url{https://swift.gsfc.nasa.gov/results/batgrbcat/index_tables.html}}. To calculate the extinction $A_r$, we use the relationship $A_V=3.1\times$$ E(B-V)$ and \citep{Schlegel1998} and $A_r/A_V\approx0.9$ \citep{Gordon2003}. We illustrate the histograms of duration and redding in Figure~\ref{fig:hist2}. 
We include new COATLI photometry of GRB 180706A (see Table~\ref{tab:grb180706A}) and GRB 180812A (see Table~\ref{tab:grb180812A}). 

\begin{figure}
\centering
 \includegraphics[width=0.45\textwidth]{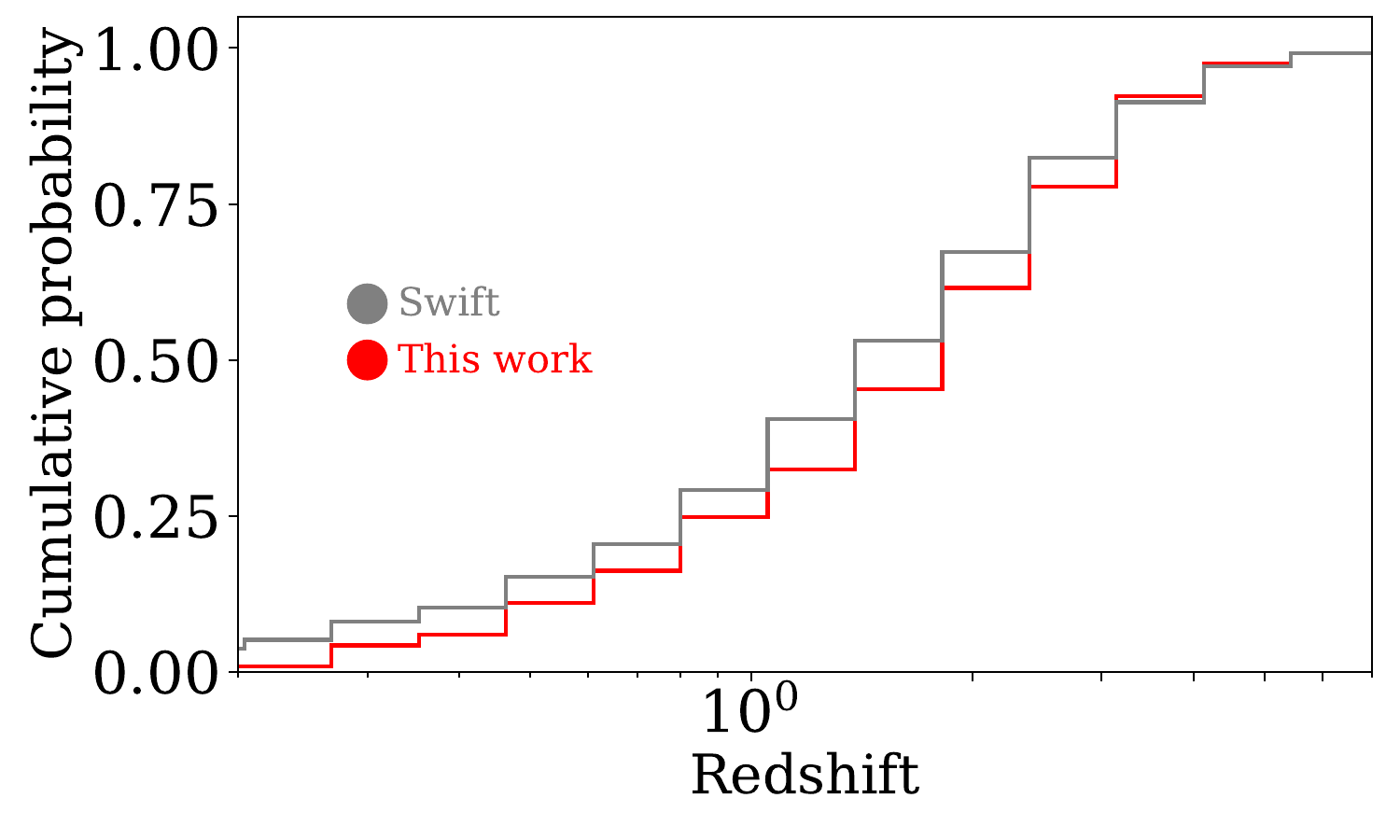}
 \caption{Comparison between the cumulative probability distribution of redshifts for the events in the sample used in this work and the redshifts of all GRBs in the {\itshape Swift}/BAT catalog.}
 \label{fig:redshiftcomparison}
\end{figure}

For the sample used in this study, we have 13 SGRBs and 205 LGRBs according to the parameter $T_{90}$. Furthermore, the Swift catalog\footnote{\url{https://swift.gsfc.nasa.gov/results/batgrbcat/summary_cflux/summary_general_info/GRBlist_redshift_BAT.txt}} gives redshifts for 116 out of 227 events, that is, 51\% of the sample. To find out if this sample is representative of the total population of GRBs, we compared the cumulative distribution of the redshifts of GRBs detected in our sample (see Table~3) and the redshift values in the catalogue of all GRBs detected by the {\itshape Swift}/BAT instrument. We plot this comparison in Figure~\ref{fig:redshiftcomparison}. The similarity between the two curves was quantified by a Kolmogorov-Smirnov test, which gavea p-value of 0.69. Therefore, we conclude that the two distributions are similar and that the results obtained in this work the population studied here is not biased in redshift with respect to the whole {\itshape Swift} sample. This distribution is analysed in detail in \S~\ref{sec:redshift}.

\section{Phenomenological Results}
\label{sec:obsresults}

\begin{figure}
\centering
\begin{tabular}{cc}
\includegraphics[width=0.45\textwidth]{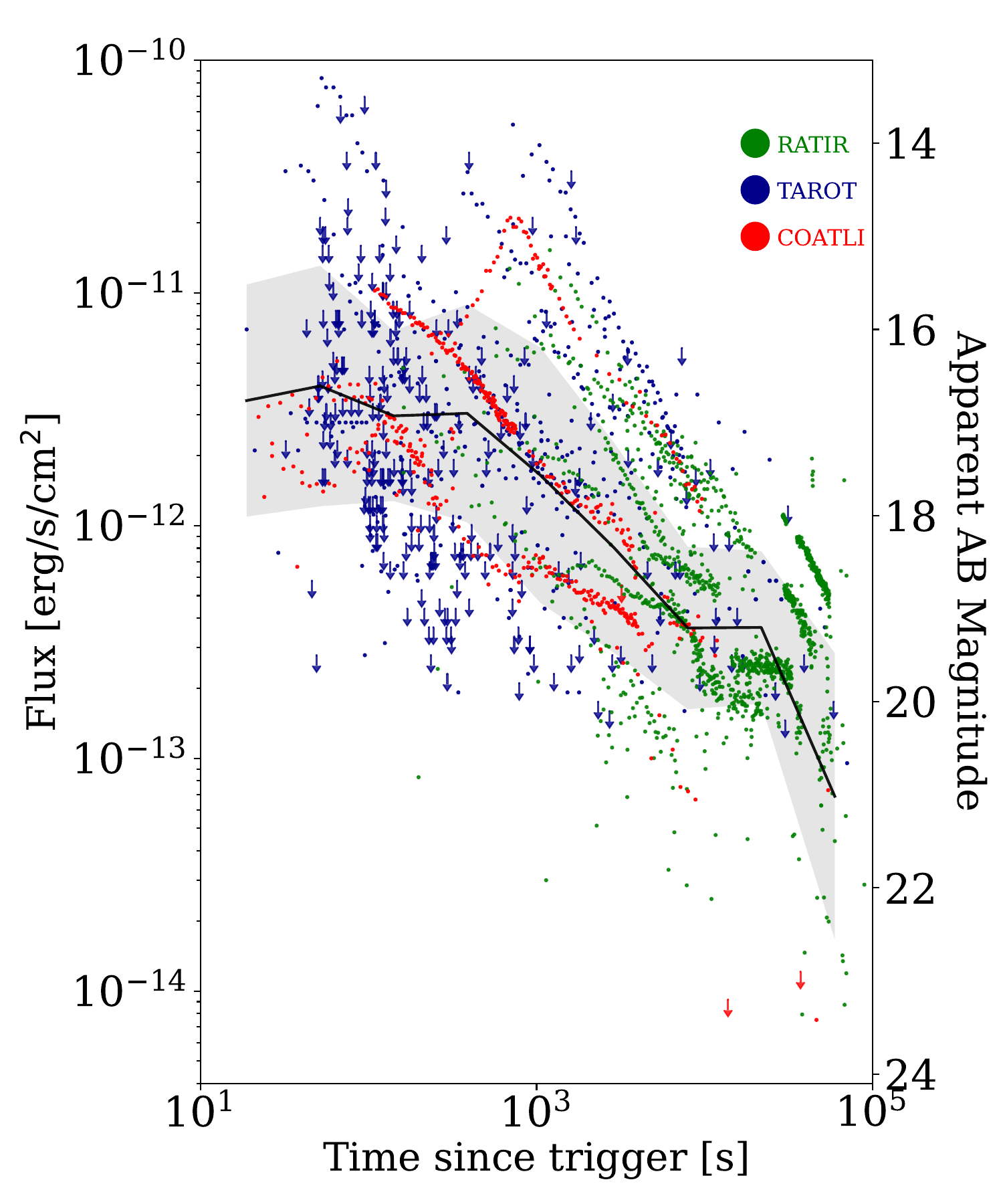}\\
\end{tabular}
 \caption{Optical detections and upper limits from TAROT (blue), RATIR (red) and COATLI (green) GRBs in the observed frame. Gray regions represent the 1$\sigma$ dispersion from the average curve (black line).}
 \label{fig:canonicalobs}
\end{figure}

A total of 227 events were collected from the COATLI, RATIR and TAROT databases, of which 133 were detections, and the remaining 94 were only upper limits, all of which are shown in Figure~\ref{fig:canonicalobs}. The first image was captured, on average, about 150~s after the trigger, with an initial AB magnitude of around $r=16.5$. 
%A total of 46 GRBs have data in such temporal interval, comprising 31 detections and 15 upper limits.
In Figure~\ref{fig:canonicalobs}, we also show the average optical light curve of prompt emission and early afterglow for GRBs (black line) with the corresponding $1\sigma$ deviation (grey regions). 

\subsection{Brightness function}

\cite{Akerlof2007} and \cite{Klotz2009} investigated the brightness function of GRBs at $T=1000$~s. We adopt a similar approach here for the bright.
We use only data between $T+500$~s and $T+2000$~s for interpolation (31 detections and 15 upper limits),compared to \cite{Akerlof2007} who used the photometric information between $T+100$~s and $T+10000$~s, we refined our fit .
For the cases where only a single data point (either detection or upper limit) was available, we estimate the magnitude  at $T=1000$s using a power-law function $F\propto t^{-\alpha}$, with $\alpha=1$ \citep{Gehrels2004}. In the case where multiple upper limits were available for the same event, we use the deepest upper limit for the calculation.

%Figure~\ref{fig:luminosity} presents the normalised cumulative probability histogram resulting from our analysis. In comparison, we use the data sample from \cite{Akerlof2007} (see their Tables~1 and 2) considering their detections and upper limits brighter than $r=18.5$, which is the limiting magnitude of COATLI and TAROT with stacked images. Due to this limitation and the difference with the limiting magnitudes of the facilities involved in the analysis performed by \cite{Akerlof2007}, we decided to set the value at $r=18.5$ in Figure~\ref{fig:luminosity}.

\begin{figure}
\centering
\includegraphics[width=0.45\textwidth]{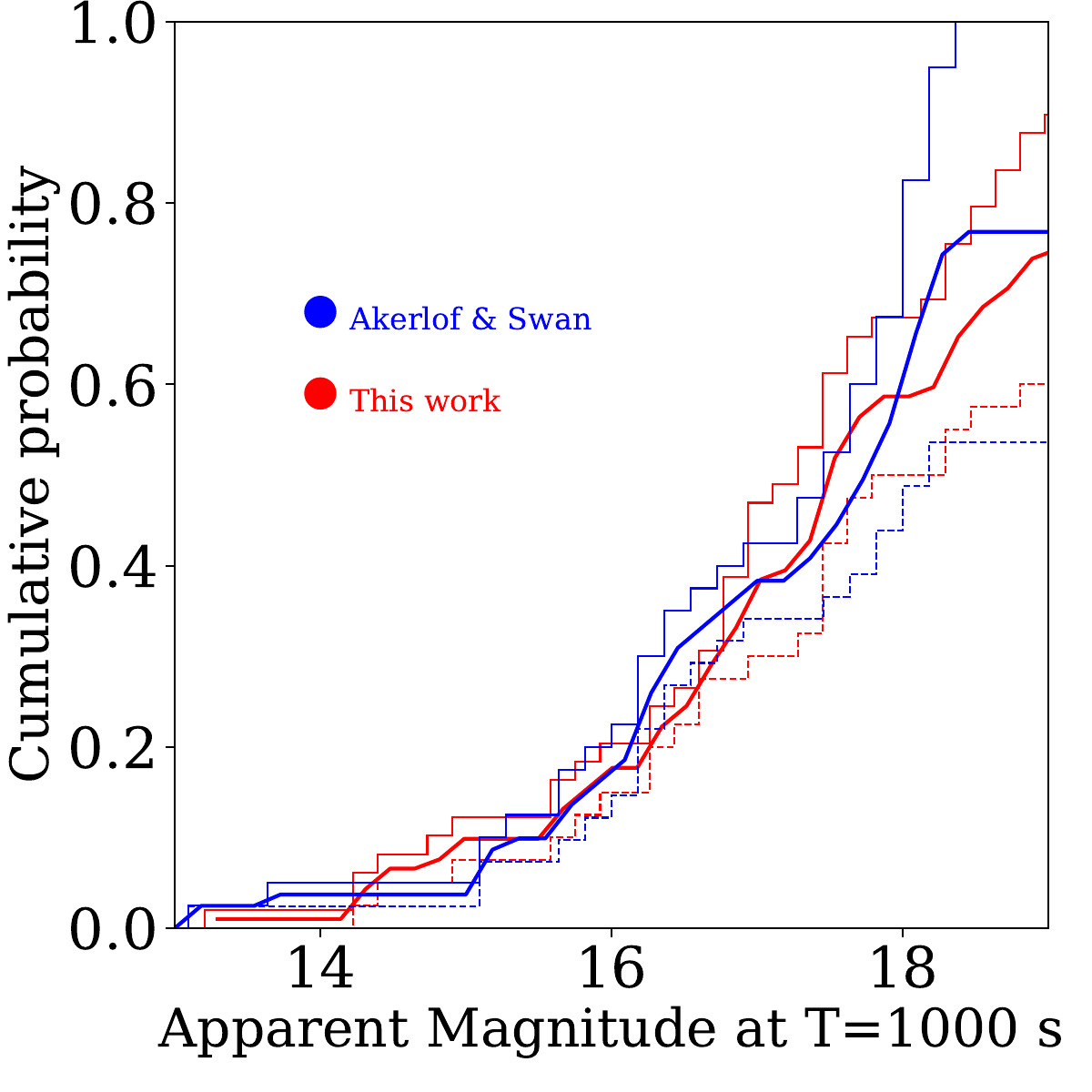}
\caption{Cumulative optical brightness function obtained in this work at $T+1000$~seconds for the detections and upper limits. We used the same criteria for pessimistic (dashed lines) and optimistic (solid lines) scenarios described by \citet{Klotz2009} for this work \citet{Akerlof2007} (blue). We also plot the mean distribution, for both samples in thicker solid lines.}
\label{fig:luminosity}
\end{figure}

Figure~\ref{fig:luminosity} presents the normalised cumulative probability histogram resulting from our analysis. We compare against a similar analysis performed with the data presented by \cite{Akerlof2007} (see their Tables~1 and 2) . However, to account for the difference with the limiting magnitudes of the facilities involved in our and their analyses, we only consider their data when the detections or upper limits are brighter than $r=18.5$, which is the limiting magnitude of COATLI and TAROT with stacked images.

Following \cite{Klotz2009}, we considered two scenarios for the upper limits: i) an optimistic scenario, in which we assume that the non-detections lie just below the observed limiting magnitude (continuous lines in Figure~\ref{fig:luminosity}); ii) a pessimistic scenario, in which we assume that the non-detections are even fainter than the faintest detected afterglow (dashed lines in Figure~\ref{fig:luminosity}). 
We also plot the average of these two scenarios (continuous thick lines in Figure~\ref{fig:luminosity}). 

Under these assumptions, by comparing our result with that obtained by \cite{Akerlof2007} we see a striking similarity in the brightness distribution. We estimate that about 20\% of the GRBs observed with telescopes such as COATLI or TAROT exhibit optical magnitude $r<$16 and 75\% $r<$18.

\subsection{Slope distribution}
\label{sec:slope}

\begin{figure*}
\centering
 \includegraphics[width=0.45\textwidth]{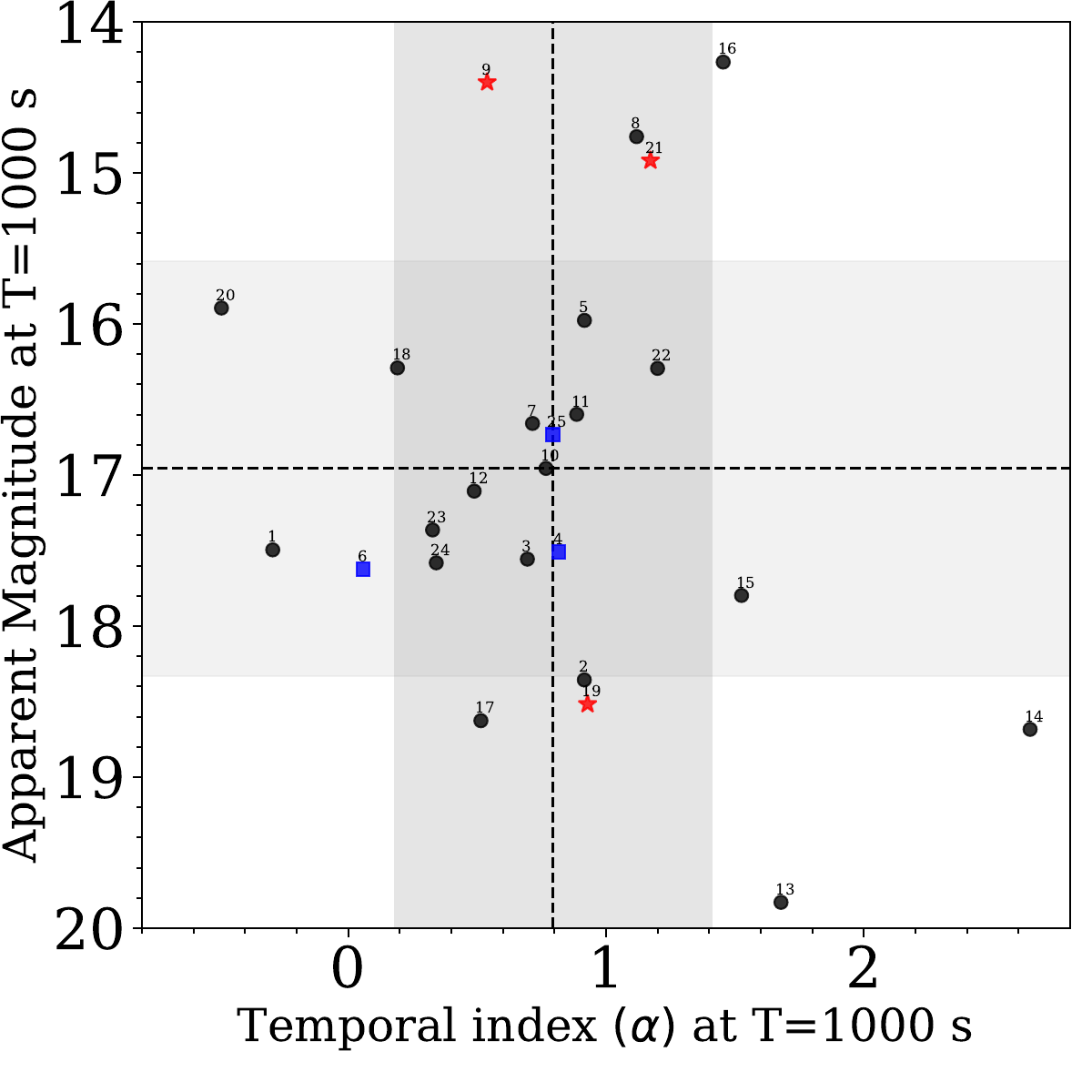}
  \includegraphics[width=0.45\textwidth]{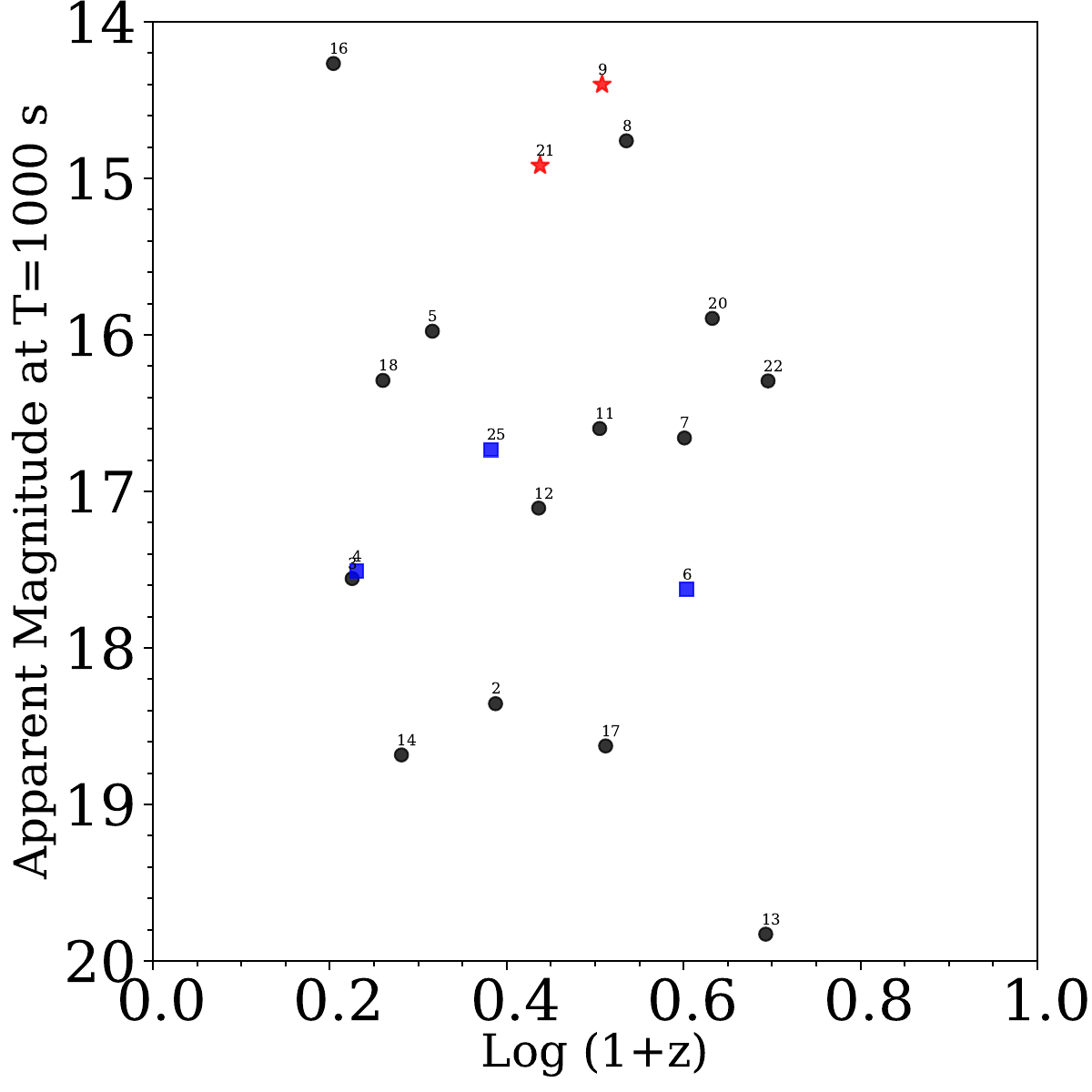}
 \caption{Left: Apparent optical magnitude vs temporal index $\alpha$ at $T+1000$~s. We plot the mean values for the apparent magnitude and the temporal index at that time (dashed black lines) and the $1\sigma$ dispersion (grey area). Right: Apparent optical magnitude at $T+1000$~s vs redshift. We distinguish the events with previously identified reverse shocks (red stars) or late central activity (blue squares). The numbers in the figure identify each GRB: 1. GRB 160127A, 2. GRB 180314A,
3. GRB 111209A, 4. GRB 060904B, 5. GRB 201021C, 6. GRB 070420, 7. GRB 141109A,
8. GRB 080413A, 9. GRB 110205A, 10. GRB 120913B, 11. GRB 210222B, 12. GRB 120119A, 13. GRB 120909A, 14. GRB 141225A, 15. GRB 130605A, 16. GRB 130215A, 17. GRB 180325A, 18. GRB 170519A, 19. GRB 180418A, 20. GRB 191016A, 21. GRB 210822A, 22. GRB 140419A, 23. GRB 150811A, 24. GRB 180812A, 25. GRB 180205A.}
\label{fig:slopes}
\end{figure*}

To gain further insight into the nature of the brightness function, we analyse the temporal index of the light curve determined for GRBs with more than one single point detected between $T+500$~s and $T+2000$~s.
We fit a single power-law function to find the temporal index and the apparent magnitude at $T=1000$~s.
We illustrate this in the left panel of Figure~\ref{fig:slopes}. We found that the average $\alpha$ for detections is $0.80 \pm 0.61$ whereas the average apparent magnitude at 1000~s is $16.72\pm 1.36$. The uncertanities refer to the $1\sigma$ deviation from the average. The value of the temporal index is slightly larger than the one reported by \cite{Dainotti2022} and \cite{Srinivasaragavan2020} (at the end of the plateau phase). This is explained by the presence of flares, reverse shocks and optical rebrightenings that can be found at earlier times.
%We use this figure to verify if the the faintest afterglows are also the fastest decaying ones. 
Figure ~\ref{fig:slopes} shows no clear correlation, indicating that the afterglow decay rate is not a dominant parameter in the brightness of the early afterglow.
We also determined in which phase of the light curve the GRBs are at time $T+1000$~s,  in order to distinguish the origin of the brightness, i.e., if the emission is produced by the typical decay of an afterglow or instead is due to an additional component (reverse shock, late central activity, etc.). 
For subsequent analysis and discussion, we distinguish events with previously identified reverse shock (stars) or late central activity components (squares).% We show the different phases recognised in the light curves, sometimes with rapid decay ($\alpha>2$) or with early rise ($\alpha<0$). 
The figure shows that there is no clear correlation between the phase of the light curve and and the observed magnitude. 

The next step was to investigate the correlation between the apparent magnitude and the distance of the GRB. This is shown in the right panel of Figure~\ref{fig:slopes}. As might be expected, a slight trend is observed between the faintest and most distant GRBs. Nevertheless, the dispersion in the graph is too large to find a conclusive empirical relationship.

\subsection{Dark GRBs}
\label{sec:dark}
Since the first detections of the afterglows in the 1990s, it has been evident that X-rays and optical light curves do not follow a canonical pattern or exhibit a standard relationship. The GRBs which are bright in X-rays and subluminous in optical, or even completely lacking the optical counterpart are called \emph{dark GRBs.}
\cite{Fynbo2001,Lazzati2002} showed that a large fraction, about 60\%-70\%, of well-localised GRBs, has detections at optical wavelengths.

\citet{Lang2010} discussed three scenarios in which the optical emission is not present or is very weak compared to typical events: i) absorption by the host galaxy or the surrounding gas \citep[e.g. see][]{Djorgovski2001,Fynbo2001}, ii) galactic absorption due to a high redshift, and iii) intrinsic faintness of some GRBs.

While instruments such as COATLI (and even more TAROT) do not have the sensitivity to detect dark GRBs, they can constrain the X-ray-to-optical flux ratio at early times. The comparison of this ratio with the bright-dark limit defined by \cite{Jakobsson2004} may indicate if early measurements with instruments like COALI or TAROT can quickly identify dark GRBs that require specific follow-up.

\cite{Jakobsson2004} proposed a definition of dark bursts based on the optical-to-X-ray spectral index $\beta_\mathrm{ox}<0.5$. This criterion has the advantage over others of leaving aside many of the factors, such as the collimation of the outflow and its density and depends on distance-independent properties of the burst rather than distance-dependent ones \citep[see e.g.][]{Djorgovski2001} and only considers how the optical counterpart is compared to the fireball model.

\begin{figure*}
\centering
 \includegraphics[width=0.8\textwidth]{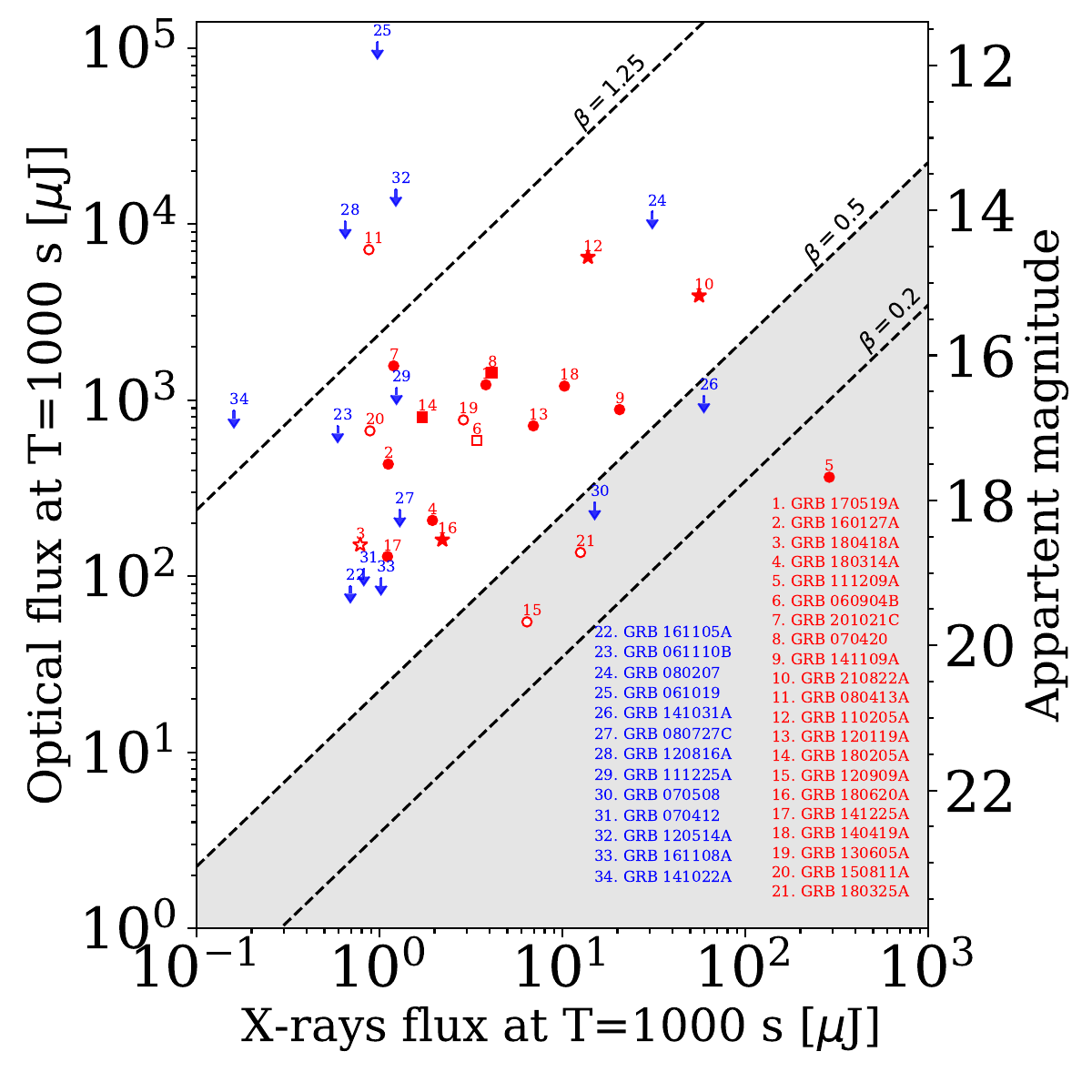}
\caption{X-ray and optical fluxes at $T+1000$~s. We show detections (red) and the upper limits (blue). The empty markers refer to events whose interpolation in the X-ray time range is uncertain. We illustrate the criteria proposed by \citet{Jakobsson2004} to identify dark GRBs (dashed black lines) and the dark GRBs region (grey area). The dashed lines, in particular, correspond to different values of the optical-to-X-ray spectral index $\beta$. }
 \label{fig:dark}
\end{figure*}

We used this criterion for our sample. Using the {\itshape Swift}/XRT light curve repository \citep{Evans2007,Evans2009} hosted at the UK {\itshape Swift} Science Data Centre (UK SSDC), we selected only GRBs for which the X-ray data had a signal-to-noise ratio $S/N>3$ at $T=1000$~s. This cuts the sample from 227 to 34.   For these, our optical photometry gives 21 detections and 13 upper limits at $T+1000$~s. These data are shown in Figure~\ref{fig:dark}. We also identified the better constraint events (with an evident temporal evolution) with filled markers and suspicious ones with unfilled markers, whose behaviour in the interpolation is ambiguous because of the presence of a plateau, reverse shock, flare, etc. We also plot different dashed lines showing the corresponding spectral index between the optical and X-rays. The \emph{dark} region, in which $\beta_\mathrm{ox}<0.5$, is shown in grey in the figure.

Figure~\ref{fig:dark} shows that all but one of the 34 GRBs lie above the bright-dark boundary, showing that the early identification of dark GRBs require instruments that are more sensitive than COATLI or TAROT. This will, however, become possible with COLIBRÍ, which reaches a limiting magnitude $R = 22$ in 1000~s \citep{Basa2022}.

In our sample, the distance (and therefore the hydrogen absorption in the optical bands due to high redshifts) is not a problem. Likewise, only one of the GRBs in the sample (the ultra long GRB~111205A) is indisputably dark under this criterion (whereas the upper limits for GRB~141031A and GRB~070508 make them likely candidates). This GRB shows a very different duration and energy-fluence than any other observed \citep{Gendre2013,Greiner2015}. Therefore, it supports the idea that is had a different origin from the populations that produce the common long and short GRBs \citep{Levan2014,Gendre2013, Greiner2015}.

\section{Physical Results}
\label{sec:phyresults}

\subsection{Redshift distribution}
\label{sec:redshift}

We calculate the redshift distribution for those GRBS in our sample that have redshift determinations following \cite{Porciani2001}. The number density of GRBs at redshift \emph{z} about \emph{z+dz} is given by:

\begin{equation}
\label{eq:number}
n(z)=\frac{dN}{dz}=\frac{R_\mathrm{GRB}(z)}{1+z}\frac{dV(z)}{dz}
\end{equation}

where $R_\mathrm{GRB}(z)$ is the comoving GRB rate as a function of \emph{z}, the factor $1/(1+z)$ accounts for the cosmological time dilation of the observed rate, and $\frac{dV(z)}{dz}$ is the comoving volume element. Assuming the star metallicity history from \cite{Kistler2008,Li2008,Virgili2011},

\begin{equation}
\label{eq:3}
R_\mathrm{GRB}(z)=\rho_0R_\mathrm{SFR}(z){(1 +z)}^{\delta}\Theta(\epsilon,z)
\end{equation}

where $\rho_0$ is the local GRB rate (in units of Gpc$^{-3}$yr$^{-1}$), $(1+z)^\delta$ accounts for the possible effect of evolution of the GRB rate exceeding the SFR rate, $\Theta(\epsilon,z)$ is the fractional mass density belonging to the metallicity below $\epsilon (Z_\odot)$ at a given z and $(Z_\odot)$ refers to the solar metal abundance). We use the star-forming rate $R_\mathrm{SFR}(z)$ reported by \cite{Porciani2001} and \cite{Guetta2007}.

\begin{equation}
{R_\mathrm{SFR}(z) =\rho_0\frac{23 \exp{(3.4z)}}{\exp{(3.4z)} + 22}}
\end{equation}

$\Theta(\epsilon,z)$ is parameterized as \cite{Hopkins2006,Langer2006} 

\begin{equation}
\Theta(\epsilon,z)=\frac{\widehat{\Gamma}(\alpha+2,\epsilon^{\beta}10^{0.15\beta z})}{\Gamma(\alpha+2)}
\end{equation}

\begin{figure}
\centering
 \includegraphics[width=0.45\textwidth]{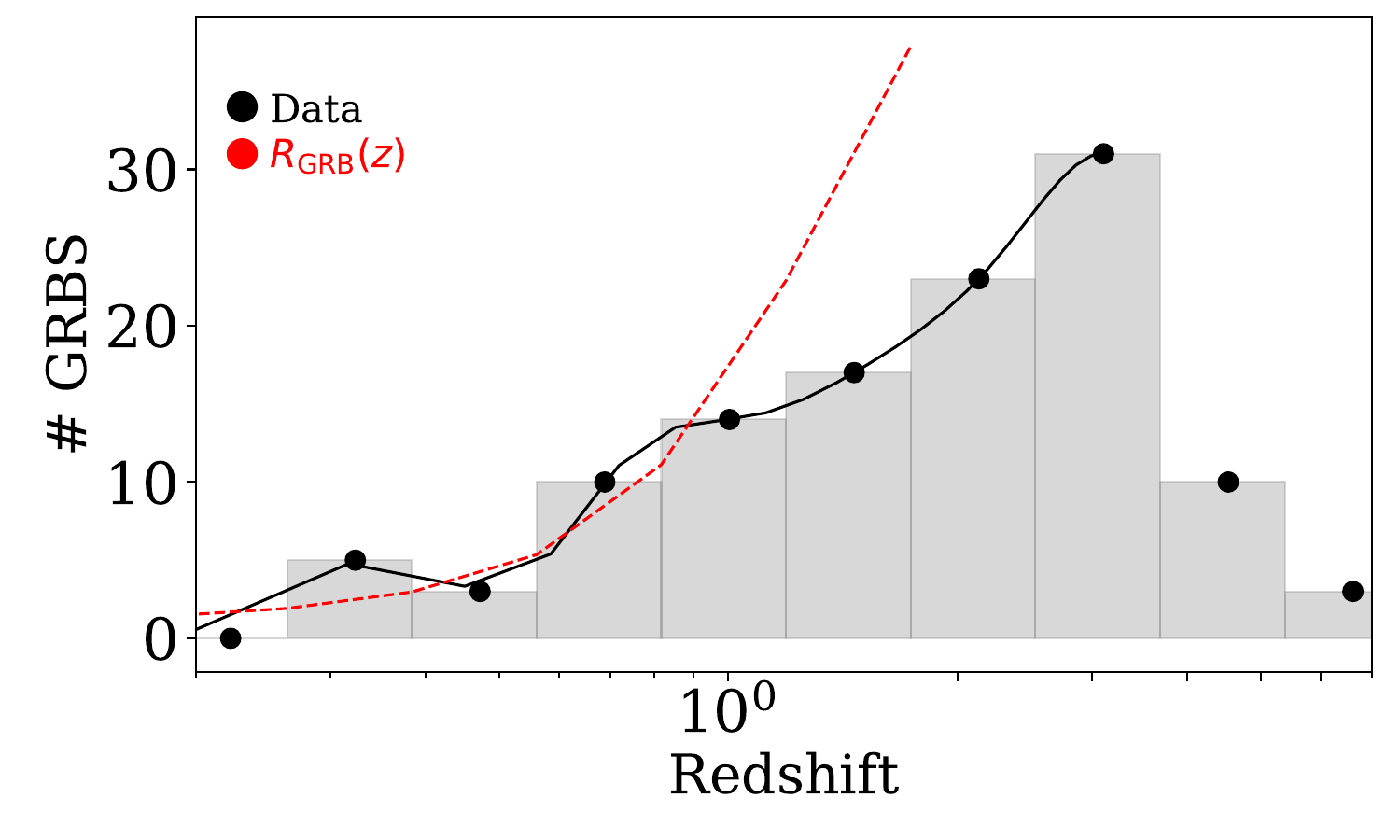}
 \caption{Histogram of the redshift distribution of our sample (gray bars) with the median data of each bin (black points) and a smoothing function for better visualization (black line). We fit a  $R_\mathrm{GRB}(z)$ function (see in \S~\ref{sec:redshift}) obtaining a $\rho_0=0.2$ Gpc$^{-3}$yr$^{-1}$ (dashed red line) with a $\chi^2=2.63$ for GRBs with redshift $z<1$. The information of redshifts was obtained from the {\itshape Swift} catalog.}
 \label{fig:redshift}
\end{figure}

where $\alpha=~-1.16$ is the power law index in the Schechter distribution function of galaxy stellar masses \citep{Panter2004} and $\beta=~2$ is the slope in the linear bisector fit to the galaxy stellar mass-metallicity relation \citep{Savaglio2006,Li2008}, $\widehat{\Gamma}$ and $\Gamma$ the incomplete and complete gamma function, respectively, $\epsilon=0.4$, and $\delta=0.4$ \citep{Li2008, Modjaz2008}.

We fit equation~\ref{eq:3} using TAROT, RATIR and COATLI data and the values/functions described previously. We estimate a GRB rate of  $\rho_0=0.2$ Gpc$^{-3}$yr$^{-1}$ (dashed red line in Figure~\ref{fig:redshift}) with a $\chi^2=2.63$ for GRB with redshift $z<1$. This is only about 40\% of the total number of GRBs expected to be detected per year $\rho_0=0.5-1.0$ \citep{Zhang2018}.

Previous studies have estimated the redshift-luminosity distribution of observed long {\itshape Swift} GRBs to obtain their rate and luminosity function. \citet{Wanderman2010} described their sample with an increase for $0<z<3$ and a decrease for $z>3$. The same qualitative behaviour was obtained by \citet{Petrosian2015} using a non-parametric determination for 200 {\itshape Swift} long GRBs with known redshifts and by \citet{Yu2015} using Monte Carlo simulations.

%This is only a fraction of the total number of GRBs expected to be detected per year $\rho_0=0.5-1.0$ \citep{Zhang2018}. however, there is a bias in the sensitivity and distance of the telescopes/instrument used in this study (see the right panel of Figure~\ref{fig:slopes}). Previous studies estimate the rate using only X-ray samples. Unfortunately, COATLI and TAROT are designed to observe the early afterglow, and therefore this study has a major limitation there. However, comparing the calculated rate, we estimate an observation of about 40\% of the total GRBs. 

Our results are consistent with the fact of COATLI and TAROT can only observe a fraction of GRBs because of factors such as weather, maintenance, and observatory closures. Furthermore, we sometimes do not observe GRBs that have previously reported magnitudes or upper limits that are below our sensitivity.

Our estimation illustrates the difficulties present in these predictions based on biased samples with uncertain selection effects, as has been discussed by \citet{Dainotti2021,Dainotti2015}.

\subsection{Canonical optical light curve}

For each GRB in our sample with a redshift determination, we produced a rest-frame light curve, averaging the photometric information for each event over 20 equally spaced logarithmic intervals between the minimum and maximum values (see Figure~\ref{fig:canonicalobs}). We corrected for Galactic extinction. To account for cosmological effects in our sample, we use the Cosmology calculator of Ned Wright\footnote{\url{http://www.astro.ucla.edu/~wright/CC.python}} \citep{Wright2006}. We assume a $\Lambda$CDM model with a $H_0=67.8$ $\mathrm{km\ s^{-1}\ Mpc^{-1}}$ \citep{Planck2014}. 

\begin{figure}
\centering
\begin{tabular}{cc}
\includegraphics[width=0.45\textwidth]{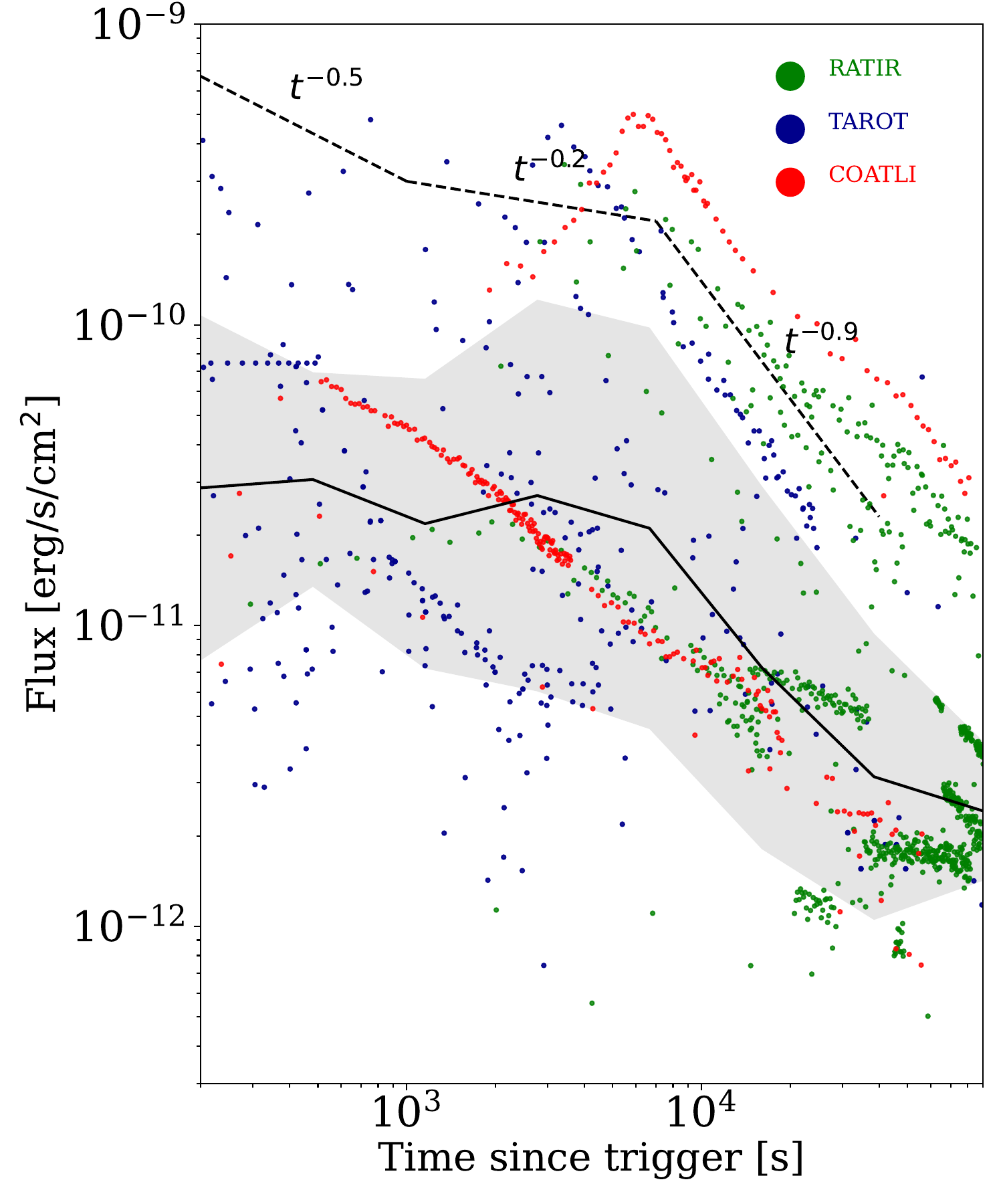}\\
\end{tabular}
 \caption{
 Optical detections from TAROT (blue), RATIR (green) and COATLI (red) GRBs in the rest frame and corrected for Galactic extinction. Gray regions represent the 1$\sigma$ dispersion about the average curve (black line). We also plot the canonical behaviour and the corresponding temporal indices (dashed line) estimated in this work.}
 \label{fig:canonicalrest}
\end{figure}

Figure~\ref{fig:canonicalrest} shows the rest-frame light curves. It also shows the average light curve as a solid line and the $1\sigma$ dispersion about this average. the $1\sigma$ error area in grey regions. 

 \cite{Zhang2006} studies the X-ray light curves of GRBs and identified the different stages described in \S~\ref{sec:introduction}.
According to \cite{Zhang2006}, the first part of the X-ray curve is likely related to tail emission of the prompt phase. After this initial decay, there is a plateau phase decay between $T=200$~s and $T=5000$~s.
%(see left panel of Figure~\ref{fig:canonicalrest}).
This segment suggests the presence of a continuing power source for the forward shock, which keeps being refreshed for some time. The most popular scenario is late central activity, that is, reactivation of the central engine. Nevertheless, it is not possible to reject other models such as a wide distribution of the shell Lorentz factors or the deceleration of a Poynting flux-dominated flow. That said, the presence of X-ray flares, around $T+1000$~s, suggests that the central engine is the source responsible for this continuing energy supply \citep{Zhang2006}. For $T>10000$, we observe the theoretically predicted normal decay.

Similarly, the optical light curve obtained in this study can be interpreted as follows:

\begin{itemize}
  \item $T<1000$ seconds: Beginning of the optical emission. Compared with x-rays, the early times of the optical emission suggests a different origin for this phase and is therefore not linked to the end of the prompt phase or internal shocks. We observe the signature of a shallow decay with an average temporal index of $\alpha=-0.5$ for $T<1000$. This has been proposed to be the combination of reverse and forward shocks, such as in the cases of GRB 180418A \citep{Becerra2019b}, GRB 180620A \citep{Becerra2019c}  GRB 060904B, GRB 070420 \citep{Klotz2008} GRB 110205 \citep{Gendre2013,Steele2017}. 
  
  \item $1000<T<7000$~s: We find a shallow temporal decay with $\alpha=-0.2$ in this interval. Nevertheless, the large dispersion illustrates the variety of features present. This might have its origin in the energy released by late central engine activity that produces a plateau phase or an optical flare \citep[see e.g.][]{Kumar2015,Zhang2018,Becerra2019a,Pereyra2022}. 
  
  \item $T>7000$~s: Optical counterpart of the normal decay observed in the X-ray canonical light curve, which is expected from the interaction between the head of the jet and the circumstellar medium. This has an average temporal index of $\alpha=-0.9$ \citep{Kumar2015,Zhang2018,Becerra2019c}.
\end{itemize}

The phases identified here show the variety of processes and components that may be involved in early optical emission.

\subsection{Parameters of the fireball model}
\label{subsec:fireball}

The fireball model is typically used to describe the general behaviour of a GRB \citep{Sari1998}. This model relies on physical parameters of the jet such as energy $E_0$, angle of aperture $\theta_\mathrm{core}$, electron index $p$, and thermal energy fractions in electrons $\epsilon_e$ and in the magnetic field $\epsilon_B$ \citep{Sari1998}. Additionally, the density $n_0$ of the medium surrounding the progenitor must also be considered \citep{Sari1998,Granot2002}. 

\begin{figure}
\centering
 \includegraphics[width=0.45\textwidth]{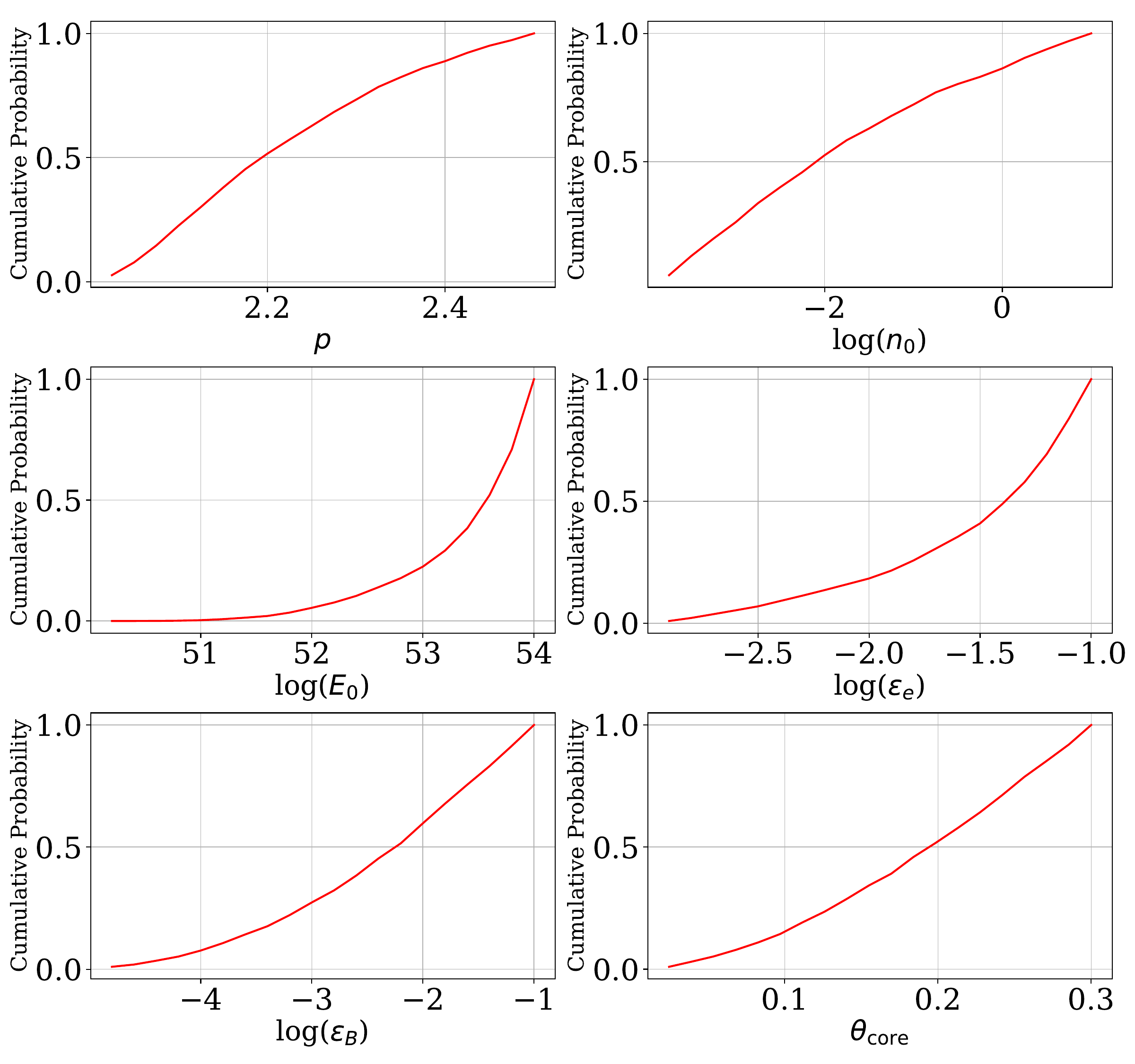}
 \caption{Cummultive distribution function (CDF) of the parameters obtained by simultaneously fitting the average X-ray and optical light curves (shown in Figure~\ref{fig:canonicalrest}) using the {\sc AFTERGLOWPY} \citep{Ryan2020} library.}
 \label{fig:histparameters}
\end{figure}

These variables are often degenerate and there are numerous combinations that could produce a given light curve.
To constrain the physics of the jet, we used the library {\sc Afterglowpy} \citep{Ryan2020} to generate 100,000 different models with uniform random distributions of the parameters mentioned above. We assumed a structured jet for all the models \citep{Urrutia2021,OConnor2023} and compare with synthetic light curves that simultaneously deviated no more than $2\sigma$ from the average X-ray and optical light curves (see Figure~\ref{fig:canonicalrest}) in the range between $T+200$~s and $T+50000$~s. The cumulative distribution functions of these constrained parameters are presented in Figure~\ref{fig:histparameters}. 

We determined the median values (corresponding to a cumulative probability of 0.5 in Figure~\ref{fig:histparameters}) for these parameters as follows: 
the energy $E_{0}\approx 10^{53.6}$~erg, an opening angle $\theta_\mathrm{core}\sim 0.2$~rad, the density $n_\mathrm{0}\sim 10^{-2.1}$ cm$^{-3}$, 
$\epsilon_e\sim 10^{-1.37}$, magnetic field $\epsilon_B\sim10^{-2.26}$, and the index of electrons $p\sim 2.2$, which are typical values for GRBs \citep[see e.g.][]{Fong2015,Santana2014,Berger2014,Levan2016} but the $\epsilon_B$ which is quite larger than previously measured \citep{Santana2014}. 

 \section{Discussion and summary}
\label{sec:discussion}

We presented optical photometry of the prompt emission and afterglow of GRBs that were observed by TAROT, COATLI, and RATIR up to 2022. We obtained an average light curve in both the observer frame and the rest frame.
We divided the results into two main categories: phenomenological and physical.

For the first part, a brightness distribution of our sample is calculated and compared to the values reported by \cite{Akerlof2007,Klotz2009} at $T+1000$~s. Nevertheless, the number of GRBs with reverse shock components or rebrightenings caused by late central activity makes it necessary to understand the behaviour of light curves beyond a simple decay. Therefore, we analyse the behaviour of the temporal indices for each GRB at this time. We observe a typical temporal index of $\alpha=0.80\pm 0.61$ for most of GRBs, but there are some cases where at that time a steeper decay (GRB 141225A) or a rapid rise (GRB 160127A and GRB 191016A) has been observed.

To investigate the nature of dark GRBs, we used the criterion proposed by \cite{Jakobsson2004}, comparing the X-rays (from the {\itshape Swift}/XRT database) and optical fluxes at $T+1000$~s to evaluate how many of the optical fluxes correspond to those predicted under the fireball model.

Physically, using an analysis of the redshift distribution of our sample, we found a local GRB rate of $\rho_0=0.2$ Gpc$^{-3}$yr$^{-1}$ for GRBs with redshift $z<1$. Therefore, compared to the lowest theoretical predictions, these collaborations were only able to observe about 40\% of the nearest GRBs. 

Additionally, using the {\sc Afterglowpy} library, we have determined a set of parameters (energy, opening angle, etc.) that deviate by a maximum of $2\sigma$ from the average canonical curve in X-rays and optical frequencies.

Although in this analysis we have worked with GRBs with cosmological and extinction corrections, this sample is limited by the instrumental characteristics of the instruments involved. For example, using photometry from other networks such as the Rapid Telescopes for Optical Response (RAPTOR) \citep{Vestrand2002}, the Mobile Astronomical System of TElescope Robots (MASTER) \citep{Lipunov2010,Lipunov2022}, the Katzman Automatic Imaging Telescope (KAIT) \citep{Li2003} and the Burst Observer and Optical Transient Exploring System (BOOTES) \citep{Castro-Tirado1999}, 
we might have better coverage in terms of follow-up. Furthermore, given the sensitivities of each of these instruments, the databases could complement each other very well. In the future, we plan to expand this study using data available from these collaborations.
Moreover, the community will also have the possibility to expand the understanding of earlier stages of GRBs with the arrival of the COLIBR\'I telescope\footnote{\url{https://colibri.lam.fr}} \citep{FuentesFernandez2020,Basa2022,Nouvel2022,Nouvel2023}. This optical-infrared telescope will offer significantly improved sensitivity compared to COATLI or TAROT but a similarly fast response. This will enable the estimation of GRB's redshift within minutes, making the upcoming years crucial for advancing our understanding of the GRB phenomenon.

\section*{acknowledgments}

We thank our anonymous referee for comments that helped us significantly improve the discussion.

TAROT has been built with the support of the Institut National des Sciences de l'Univers, CNRS, France. TAROT is funded by the CNES and thanks to the help of the technical staff of the Observatoire de Haute Provence, OSU-Pytheas.

We thank the staff of the Observatorio Astron\'omico Nacional on Sierra San Pedro M\'artir. Some of the data used in this paper were acquired with the COATLI telescope and interim instrument at the Observatorio Astron\'omico Nacional on the Sierra de San Pedro M\'artir. COATLI is funded by CONACyT (LN 232649, 260369, and 271117) and the Universidad Nacional Aut\'onoma de M\'exico (CIC and DGAPA/PAPIIT IT102715, IG100414, IN109408, and IN105921) and is operated and maintained by the Observatorio Astron\'omico Nacional and the Instituto de Astronom\'ia of the Universidad Nacional Aut\'onoma de M\'exico.

RATIR is a collaboration between the University of California, the Universidad Nacional Auton\'oma de M\'exico, NASA Goddard Space Flight Center, and Arizona State University, benefiting from the loan of an H2RG detector and hardware and software support from Teledyne Scientific and Imaging. RATIR, the automation of the Harold L. Johnson Telescope of the Observatorio Astron\'omico Nacional on Sierra San Pedro M\'artir, and the operation of both are funded through NASA grants NNX09AH71G, NNX09AT02G, NNX10AI27G, and NNX12AE66G, CONACyT grants INFR-2009-01-122785 and CB-2008-101958, UNAM PAPIIT grants IG100414, IA102917, and IN105921, UC MEXUS-CONACyT grant CN 09-283, and the Instituto de Astronom{\'\i}a of the Universidad Nacional Auton\'oma de M\'exico. The authors acknowledge the vital contributions of Neil Gehrels and Leonid Georgiev to the early development of RATIR.

RLB acknowledges support from the CONAHCyT postdoctoral fellowship.

This work made use of data supplied by the UK {\itshape Swift} Science Data Centre at the University of Leicester.

\section*{Data availability}
The data underlying this article will be shared on reasonable request to the corresponding author.



\bibliographystyle{mnras}
\bibliography{references} 

\begin{thebibliography}{}
\makeatletter
\relax
\def\mn@urlcharsother{\let\do\@makeother \do\$\do\&\do\#\do\^\do\_\do\%\do\~}
\def\mn@doi{\begingroup\mn@urlcharsother \@ifnextchar [ {\mn@doi@}
  {\mn@doi@[]}}
\def\mn@doi@[#1]#2{\def\@tempa{#1}\ifx\@tempa\@empty \href
  {http://dx.doi.org/#2} {doi:#2}\else \href {http://dx.doi.org/#2} {#1}\fi
  \endgroup}
\def\mn@eprint#1#2{\mn@eprint@#1:#2::\@nil}
\def\mn@eprint@arXiv#1{\href {http://arxiv.org/abs/#1} {{\tt arXiv:#1}}}
\def\mn@eprint@dblp#1{\href {http://dblp.uni-trier.de/rec/bibtex/#1.xml}
  {dblp:#1}}
\def\mn@eprint@#1:#2:#3:#4\@nil{\def\@tempa {#1}\def\@tempb {#2}\def\@tempc
  {#3}\ifx \@tempc \@empty \let \@tempc \@tempb \let \@tempb \@tempa \fi \ifx
  \@tempb \@empty \def\@tempb {arXiv}\fi \@ifundefined
  {mn@eprint@\@tempb}{\@tempb:\@tempc}{\expandafter \expandafter \csname
  mn@eprint@\@tempb\endcsname \expandafter{\@tempc}}}

\bibitem[\protect\citeauthoryear{{Abbott} et~al.,}{{Abbott}
  et~al.}{2017}]{Abbott2017}
{Abbott} B.~P.,  et~al., 2017, \mn@doi [\apjl] {10.3847/2041-8213/aa920c},
  \href {https://ui.adsabs.harvard.edu/abs/2017ApJ...848L..13A} {848, L13}

\bibitem[\protect\citeauthoryear{{Ahn} et~al.,}{{Ahn} et~al.}{2012}]{Ahn2012}
{Ahn} C.~P.,  et~al., 2012, \mn@doi [\apjs] {10.1088/0067-0049/203/2/21}, \href
  {https://ui.adsabs.harvard.edu/abs/2012ApJS..203...21A} {203, 21}

\bibitem[\protect\citeauthoryear{{Akerlof} \& {Swan}}{{Akerlof} \&
  {Swan}}{2007}]{Akerlof2007}
{Akerlof} C.~W.,  {Swan} H.~F.,  2007, \mn@doi [\apj] {10.1086/523081}, \href
  {https://ui.adsabs.harvard.edu/abs/2007ApJ...671.1868A} {671, 1868}

\bibitem[\protect\citeauthoryear{{Anderson} et~al.,}{{Anderson}
  et~al.}{2018}]{Anderson2018}
{Anderson} M.~M.,  et~al., 2018, \mn@doi [\apj] {10.3847/1538-4357/aad2d7},
  \href {https://ui.adsabs.harvard.edu/abs/2018ApJ...864...22A} {864, 22}

\bibitem[\protect\citeauthoryear{{Atteia} et~al.,}{{Atteia}
  et~al.}{2017}]{Atteia2017}
{Atteia} J.~L.,  et~al., 2017, \mn@doi [\apj] {10.3847/1538-4357/aa5ffa}, \href
  {https://ui.adsabs.harvard.edu/abs/2017ApJ...837..119A} {837, 119}

\bibitem[\protect\citeauthoryear{{Basa} et~al.,}{{Basa}
  et~al.}{2022}]{Basa2022}
{Basa} S.,  et~al., 2022, in {Marshall} H.~K.,  {Spyromilio} J.,   {Usuda} T.,
  eds,  Society of Photo-Optical Instrumentation Engineers (SPIE) Conference
  Series Vol. 12182, Ground-based and Airborne Telescopes IX. p. 121821S,
  \mn@doi{10.1117/12.2627139}

\bibitem[\protect\citeauthoryear{{Becerra} et~al.,}{{Becerra}
  et~al.}{2017}]{Becerra2017}
{Becerra} R.~L.,  et~al., 2017, \mn@doi [\apj] {10.3847/1538-4357/aa610f},
  \href {https://ui.adsabs.harvard.edu/abs/2017ApJ...837..116B} {837, 116}

\bibitem[\protect\citeauthoryear{{Becerra} et~al.,}{{Becerra}
  et~al.}{2019a}]{Becerra2019a}
{Becerra} R.~L.,  et~al., 2019a, \mn@doi [\apj] {10.3847/1538-4357/ab0026},
  \href {https://ui.adsabs.harvard.edu/abs/2019ApJ...872..118B} {872, 118}

\bibitem[\protect\citeauthoryear{{Becerra} et~al.,}{{Becerra}
  et~al.}{2019b}]{Becerra2019b}
{Becerra} R.~L.,  et~al., 2019b, \mn@doi [\apj] {10.3847/1538-4357/ab275b},
  \href {https://ui.adsabs.harvard.edu/abs/2019ApJ...881...12B} {881, 12}

\bibitem[\protect\citeauthoryear{{Becerra} et~al.,}{{Becerra}
  et~al.}{2019c}]{Becerra2019c}
{Becerra} R.~L.,  et~al., 2019c, \mn@doi [\apj] {10.3847/1538-4357/ab5859},
  \href {https://ui.adsabs.harvard.edu/abs/2019ApJ...887..254B} {887, 254}

\bibitem[\protect\citeauthoryear{{Becerra} et~al.,}{{Becerra}
  et~al.}{2021}]{Becerra2021}
{Becerra} R.~L.,  et~al., 2021, \mn@doi [\apj] {10.3847/1538-4357/abcd3a},
  \href {https://ui.adsabs.harvard.edu/abs/2021ApJ...908...39B} {908, 39}

\bibitem[\protect\citeauthoryear{{Becerra} et~al.,}{{Becerra}
  et~al.}{2023}]{Becerra2023}
{Becerra} R.~L.,  et~al., 2023, \mn@doi [\mnras] {10.1093/mnras/stad1372},
  \href {https://ui.adsabs.harvard.edu/abs/2023MNRAS.522.5204B} {522, 5204}

\bibitem[\protect\citeauthoryear{{Berger}}{{Berger}}{2014}]{Berger2014}
{Berger} E.,  2014, \mn@doi [\araa] {10.1146/annurev-astro-081913-035926},
  \href {https://ui.adsabs.harvard.edu/abs/2014ARA&A..52...43B} {52, 43}

\bibitem[\protect\citeauthoryear{{Bertin}}{{Bertin}}{2010}]{Bertin2010}
{Bertin} E.,  2010, {SWarp: Resampling and Co-adding FITS Images Together},
  Astrophysics Source Code Library, record ascl:1010.068 (\mn@eprint {ascl}
  {1010.068})

\bibitem[\protect\citeauthoryear{{Bertin} \& {Arnouts}}{{Bertin} \&
  {Arnouts}}{1996}]{Bertin1996}
{Bertin} E.,  {Arnouts} S.,  1996, \mn@doi [\aaps] {10.1051/aas:1996164}, \href
  {https://ui.adsabs.harvard.edu/abs/1996A&AS..117..393B} {117, 393}

\bibitem[\protect\citeauthoryear{{Burrows} et~al.,}{{Burrows}
  et~al.}{2005}]{Burrows2005}
{Burrows} D.~N.,  et~al., 2005, \mn@doi [Science] {10.1126/science.1116168},
  \href {https://ui.adsabs.harvard.edu/abs/2005Sci...309.1833B} {309, 1833}

\bibitem[\protect\citeauthoryear{{Butler} et~al.,}{{Butler}
  et~al.}{2012}]{Butler2012}
{Butler} N.,  et~al., 2012, in {McLean} I.~S.,  {Ramsay} S.~K.,   {Takami} H.,
  eds,  Society of Photo-Optical Instrumentation Engineers (SPIE) Conference
  Series Vol. 8446, Ground-based and Airborne Instrumentation for Astronomy IV.
  p. 844610, \mn@doi{10.1117/12.926471}

\bibitem[\protect\citeauthoryear{{Castro-Tirado} et~al.,}{{Castro-Tirado}
  et~al.}{1999}]{Castro-Tirado1999}
{Castro-Tirado} A.~J.,  et~al., 1999, \mn@doi [\aaps] {10.1051/aas:1999362},
  \href {https://ui.adsabs.harvard.edu/abs/1999A&AS..138..583C} {138, 583}

\bibitem[\protect\citeauthoryear{{Cuevas} et~al.,}{{Cuevas}
  et~al.}{2016}]{Cuevas2016}
{Cuevas} S.,  et~al., 2016, in {Evans} C.~J.,  {Simard} L.,   {Takami} H.,
  eds,  Society of Photo-Optical Instrumentation Engineers (SPIE) Conference
  Series Vol. 9908, Ground-based and Airborne Instrumentation for Astronomy VI.
  p. 99085Q, \mn@doi{10.1117/12.2234200}

\bibitem[\protect\citeauthoryear{{Dainotti}, {Del Vecchio}, {Shigehiro}  \&
  {Capozziello}}{{Dainotti} et~al.}{2015}]{Dainotti2015}
{Dainotti} M.~G.,  {Del Vecchio} R.,  {Shigehiro} N.,   {Capozziello} S.,
  2015, \mn@doi [\apj] {10.1088/0004-637X/800/1/31}, \href
  {https://ui.adsabs.harvard.edu/abs/2015ApJ...800...31D} {800, 31}

\bibitem[\protect\citeauthoryear{{Dainotti}, {Petrosian}  \&
  {Bowden}}{{Dainotti} et~al.}{2021}]{Dainotti2021}
{Dainotti} M.~G.,  {Petrosian} V.,   {Bowden} L.,  2021, \mn@doi [\apjl]
  {10.3847/2041-8213/abf5e4}, \href
  {https://ui.adsabs.harvard.edu/abs/2021ApJ...914L..40D} {914, L40}

\bibitem[\protect\citeauthoryear{{Dainotti}, {Levine}, {Fraija}, {Warren}  \&
  {Sourav}}{{Dainotti} et~al.}{2022}]{Dainotti2022}
{Dainotti} M.~G.,  {Levine} D.,  {Fraija} N.,  {Warren} D.,   {Sourav} S.,
  2022, \mn@doi [\apj] {10.3847/1538-4357/ac9b11}, \href
  {https://ui.adsabs.harvard.edu/abs/2022ApJ...940..169D} {940, 169}

\bibitem[\protect\citeauthoryear{{Djorgovski}, {Frail}, {Kulkarni}, {Bloom},
  {Odewahn}  \& {Diercks}}{{Djorgovski} et~al.}{2001}]{Djorgovski2001}
{Djorgovski} S.~G.,  {Frail} D.~A.,  {Kulkarni} S.~R.,  {Bloom} J.~S.,
  {Odewahn} S.~C.,   {Diercks} A.,  2001, \mn@doi [\apj] {10.1086/323845},
  \href {https://ui.adsabs.harvard.edu/abs/2001ApJ...562..654D} {562, 654}

\bibitem[\protect\citeauthoryear{{Eichler} \& {Levinson}}{{Eichler} \&
  {Levinson}}{2000}]{Eichler2000}
{Eichler} D.,  {Levinson} A.,  2000, \mn@doi [\apj] {10.1086/308245}, \href
  {https://ui.adsabs.harvard.edu/abs/2000ApJ...529..146E} {529, 146}

\bibitem[\protect\citeauthoryear{{Evans} et~al.,}{{Evans}
  et~al.}{2007}]{Evans2007}
{Evans} P.~A.,  et~al., 2007, \mn@doi [\aap] {10.1051/0004-6361:20077530},
  \href {https://ui.adsabs.harvard.edu/abs/2007A&A...469..379E} {469, 379}

\bibitem[\protect\citeauthoryear{{Evans} et~al.,}{{Evans}
  et~al.}{2009}]{Evans2009}
{Evans} P.~A.,  et~al., 2009, \mn@doi [\mnras]
  {10.1111/j.1365-2966.2009.14913.x}, \href
  {https://ui.adsabs.harvard.edu/abs/2009MNRAS.397.1177E} {397, 1177}

\bibitem[\protect\citeauthoryear{{Fong}, {Berger}, {Margutti}  \&
  {Zauderer}}{{Fong} et~al.}{2015}]{Fong2015}
{Fong} W.,  {Berger} E.,  {Margutti} R.,   {Zauderer} B.~A.,  2015, \mn@doi
  [\apj] {10.1088/0004-637X/815/2/102}, \href
  {https://ui.adsabs.harvard.edu/abs/2015ApJ...815..102F} {815, 102}

\bibitem[\protect\citeauthoryear{{Fuentes-Fern{\'a}ndez}
  et~al.,}{{Fuentes-Fern{\'a}ndez} et~al.}{2020}]{FuentesFernandez2020}
{Fuentes-Fern{\'a}ndez} J.,  et~al., 2020, \mn@doi [Journal of Astronomical
  Instrumentation] {10.1142/S2251171720500014}, \href
  {https://ui.adsabs.harvard.edu/abs/2020JAI.....950001F} {9, 2050001}

\bibitem[\protect\citeauthoryear{{Fynbo} et~al.,}{{Fynbo}
  et~al.}{2001}]{Fynbo2001}
{Fynbo} J.~U.,  et~al., 2001, \mn@doi [\aap] {10.1051/0004-6361:20010112},
  \href {https://ui.adsabs.harvard.edu/abs/2001A&A...369..373F} {369, 373}

\bibitem[\protect\citeauthoryear{{Gehrels} et~al.,}{{Gehrels}
  et~al.}{2004}]{Gehrels2004}
{Gehrels} N.,  et~al., 2004, \mn@doi [\apj] {10.1086/422091}, \href
  {https://ui.adsabs.harvard.edu/abs/2004ApJ...611.1005G} {611, 1005}

\bibitem[\protect\citeauthoryear{{Gendre} et~al.,}{{Gendre}
  et~al.}{2013}]{Gendre2013}
{Gendre} B.,  et~al., 2013, \mn@doi [\apj] {10.1088/0004-637X/766/1/30}, \href
  {https://ui.adsabs.harvard.edu/abs/2013ApJ...766...30G} {766, 30}

\bibitem[\protect\citeauthoryear{{Gordon}, {Clayton}, {Misselt}, {Landolt}  \&
  {Wolff}}{{Gordon} et~al.}{2003}]{Gordon2003}
{Gordon} K.~D.,  {Clayton} G.~C.,  {Misselt} K.~A.,  {Landolt} A.~U.,   {Wolff}
  M.~J.,  2003, \mn@doi [\apj] {10.1086/376774}, \href
  {https://ui.adsabs.harvard.edu/abs/2003ApJ...594..279G} {594, 279}

\bibitem[\protect\citeauthoryear{{Granot}, {Panaitescu}, {Kumar}  \&
  {Woosley}}{{Granot} et~al.}{2002}]{Granot2002}
{Granot} J.,  {Panaitescu} A.,  {Kumar} P.,   {Woosley} S.~E.,  2002, \mn@doi
  [\apjl] {10.1086/340991}, \href
  {https://ui.adsabs.harvard.edu/abs/2002ApJ...570L..61G} {570, L61}

\bibitem[\protect\citeauthoryear{{Greiner} et~al.,}{{Greiner}
  et~al.}{2011}]{Greiner2011}
{Greiner} J.,  et~al., 2011, \mn@doi [\aap] {10.1051/0004-6361/201015458},
  \href {https://ui.adsabs.harvard.edu/abs/2011A&A...526A..30G} {526, A30}

\bibitem[\protect\citeauthoryear{{Greiner} et~al.,}{{Greiner}
  et~al.}{2015}]{Greiner2015}
{Greiner} J.,  et~al., 2015, \mn@doi [\nat] {10.1038/nature14579}, \href
  {https://ui.adsabs.harvard.edu/abs/2015Natur.523..189G} {523, 189}

\bibitem[\protect\citeauthoryear{{Guetta} \& {Piran}}{{Guetta} \&
  {Piran}}{2007}]{Guetta2007}
{Guetta} D.,  {Piran} T.,  2007, \mn@doi [\jcap]
  {10.1088/1475-7516/2007/07/003}, \href
  {https://ui.adsabs.harvard.edu/abs/2007JCAP...07..003G} {2007, 003}

\bibitem[\protect\citeauthoryear{{Hopkins} \& {Beacom}}{{Hopkins} \&
  {Beacom}}{2006}]{Hopkins2006}
{Hopkins} A.~M.,  {Beacom} J.~F.,  2006, \mn@doi [\apj] {10.1086/506610}, \href
  {https://ui.adsabs.harvard.edu/abs/2006ApJ...651..142H} {651, 142}

\bibitem[\protect\citeauthoryear{{Jakobsson}, {Hjorth}, {Fynbo}, {Watson},
  {Pedersen}, {Bj{\"o}rnsson}  \& {Gorosabel}}{{Jakobsson}
  et~al.}{2004}]{Jakobsson2004}
{Jakobsson} P.,  {Hjorth} J.,  {Fynbo} J.~P.~U.,  {Watson} D.,  {Pedersen} K.,
  {Bj{\"o}rnsson} G.,   {Gorosabel} J.,  2004, \mn@doi [\apjl]
  {10.1086/427089}, \href
  {https://ui.adsabs.harvard.edu/abs/2004ApJ...617L..21J} {617, L21}

\bibitem[\protect\citeauthoryear{{Kann} et~al.,}{{Kann}
  et~al.}{2010}]{Kann2010}
{Kann} D.~A.,  et~al., 2010, \mn@doi [\apj] {10.1088/0004-637X/720/2/1513},
  \href {https://ui.adsabs.harvard.edu/abs/2010ApJ...720.1513K} {720, 1513}

\bibitem[\protect\citeauthoryear{{Kistler}, {Y{\"u}ksel}, {Beacom}  \&
  {Stanek}}{{Kistler} et~al.}{2008}]{Kistler2008}
{Kistler} M.~D.,  {Y{\"u}ksel} H.,  {Beacom} J.~F.,   {Stanek} K.~Z.,  2008,
  \mn@doi [\apjl] {10.1086/527671}, \href
  {https://ui.adsabs.harvard.edu/abs/2008ApJ...673L.119K} {673, L119}

\bibitem[\protect\citeauthoryear{{Klotz}, {Gendre}, {Stratta}, {Atteia},
  {Bo{\"e}r}, {Malacrino}, {Damerdji}  \& {Behrend}}{{Klotz}
  et~al.}{2006}]{Klotz2006}
{Klotz} A.,  {Gendre} B.,  {Stratta} G.,  {Atteia} J.~L.,  {Bo{\"e}r} M.,
  {Malacrino} F.,  {Damerdji} Y.,   {Behrend} R.,  2006, \mn@doi [\aap]
  {10.1051/0004-6361:20065158}, \href
  {https://ui.adsabs.harvard.edu/abs/2006A&A...451L..39K} {451, L39}

\bibitem[\protect\citeauthoryear{{Klotz}, {Bo{\"e}r}, {Eysseric}, {Damerdji},
  {Laas{\textendash}Bourez}, {Pollas}  \& {Vachier}}{{Klotz}
  et~al.}{2008}]{Klotz2008}
{Klotz} A.,  {Bo{\"e}r} M.,  {Eysseric} J.,  {Damerdji} Y.,
  {Laas{\textendash}Bourez} M.,  {Pollas} C.,   {Vachier} F.,  2008, \mn@doi
  [\pasp] {10.1086/596022}, \href
  {https://ui.adsabs.harvard.edu/abs/2008PASP..120.1298K} {120, 1298}

\bibitem[\protect\citeauthoryear{{Klotz}, {Bo{\"e}r}, {Atteia}  \&
  {Gendre}}{{Klotz} et~al.}{2009}]{Klotz2009}
{Klotz} A.,  {Bo{\"e}r} M.,  {Atteia} J.~L.,   {Gendre} B.,  2009, \mn@doi
  [\aj] {10.1088/0004-6256/137/5/4100}, \href
  {https://ui.adsabs.harvard.edu/abs/2009AJ....137.4100K} {137, 4100}

\bibitem[\protect\citeauthoryear{{Kobayashi}, {Piran}  \& {Sari}}{{Kobayashi}
  et~al.}{1997}]{Kobayashi1997}
{Kobayashi} S.,  {Piran} T.,   {Sari} R.,  1997, \mn@doi [\apj]
  {10.1086/512791}, \href
  {https://ui.adsabs.harvard.edu/abs/1997ApJ...490...92K} {490, 92}

\bibitem[\protect\citeauthoryear{{Kouveliotou}, {Meegan}, {Fishman}, {Bhat},
  {Briggs}, {Koshut}, {Paciesas}  \& {Pendleton}}{{Kouveliotou}
  et~al.}{1993}]{Kouveliotou1993}
{Kouveliotou} C.,  {Meegan} C.~A.,  {Fishman} G.~J.,  {Bhat} N.~P.,  {Briggs}
  M.~S.,  {Koshut} T.~M.,  {Paciesas} W.~S.,   {Pendleton} G.~N.,  1993,
  \mn@doi [\apjl] {10.1086/186969}, \href
  {https://ui.adsabs.harvard.edu/abs/1993ApJ...413L.101K} {413, L101}

\bibitem[\protect\citeauthoryear{{Kumar} \& {Zhang}}{{Kumar} \&
  {Zhang}}{2015}]{Kumar2015}
{Kumar} P.,  {Zhang} B.,  2015, \mn@doi [\physrep]
  {10.1016/j.physrep.2014.09.008}, \href
  {https://ui.adsabs.harvard.edu/abs/2015PhR...561....1K} {561, 1}

\bibitem[\protect\citeauthoryear{{Lang}, {Hogg}, {Mierle}, {Blanton}  \&
  {Roweis}}{{Lang} et~al.}{2010}]{Lang2010}
{Lang} D.,  {Hogg} D.~W.,  {Mierle} K.,  {Blanton} M.,   {Roweis} S.,  2010,
  \mn@doi [\aj] {10.1088/0004-6256/139/5/1782}, \href
  {https://ui.adsabs.harvard.edu/abs/2010AJ....139.1782L} {139, 1782}

\bibitem[\protect\citeauthoryear{{Langer} \& {Norman}}{{Langer} \&
  {Norman}}{2006}]{Langer2006}
{Langer} N.,  {Norman} C.~A.,  2006, \mn@doi [\apjl] {10.1086/500363}, \href
  {https://ui.adsabs.harvard.edu/abs/2006ApJ...638L..63L} {638, L63}

\bibitem[\protect\citeauthoryear{{Lazzati}, {Covino}  \&
  {Ghisellini}}{{Lazzati} et~al.}{2002}]{Lazzati2002}
{Lazzati} D.,  {Covino} S.,   {Ghisellini} G.,  2002, \mn@doi [\mnras]
  {10.1046/j.1365-8711.2002.05076.x}, \href
  {https://ui.adsabs.harvard.edu/abs/2002MNRAS.330..583L} {330, 583}

\bibitem[\protect\citeauthoryear{{Lee} \& {Ramirez-Ruiz}}{{Lee} \&
  {Ramirez-Ruiz}}{2007}]{Lee2007}
{Lee} W.~H.,  {Ramirez-Ruiz} E.,  2007, \mn@doi [New Journal of Physics]
  {10.1088/1367-2630/9/1/017}, \href
  {https://ui.adsabs.harvard.edu/abs/2007NJPh....9...17L} {9, 17}

\bibitem[\protect\citeauthoryear{{Levan} et~al.,}{{Levan}
  et~al.}{2014}]{Levan2014}
{Levan} A.~J.,  et~al., 2014, \mn@doi [\apj] {10.1088/0004-637X/781/1/13},
  \href {https://ui.adsabs.harvard.edu/abs/2014ApJ...781...13L} {781, 13}

\bibitem[\protect\citeauthoryear{{Levan}, {Crowther}, {de Grijs}, {Langer},
  {Xu}  \& {Yoon}}{{Levan} et~al.}{2016}]{Levan2016}
{Levan} A.,  {Crowther} P.,  {de Grijs} R.,  {Langer} N.,  {Xu} D.,   {Yoon}
  S.-C.,  2016, \mn@doi [\ssr] {10.1007/s11214-016-0312-x}, \href
  {https://ui.adsabs.harvard.edu/abs/2016SSRv..202...33L} {202, 33}

\bibitem[\protect\citeauthoryear{{Li}}{{Li}}{2008}]{Li2008}
{Li} L.-X.,  2008, \mn@doi [\mnras] {10.1111/j.1365-2966.2008.13488.x}, \href
  {https://ui.adsabs.harvard.edu/abs/2008MNRAS.388.1487L} {388, 1487}

\bibitem[\protect\citeauthoryear{{Li}, {Filippenko}, {Chornock}  \& {Jha}}{{Li}
  et~al.}{2003}]{Li2003}
{Li} W.,  {Filippenko} A.~V.,  {Chornock} R.,   {Jha} S.,  2003, \mn@doi
  [\pasp] {10.1086/376432}, \href
  {https://ui.adsabs.harvard.edu/abs/2003PASP..115..844L} {115, 844}

\bibitem[\protect\citeauthoryear{{Li} et~al.,}{{Li} et~al.}{2012}]{Li2012}
{Li} L.,  et~al., 2012, \mn@doi [\apj] {10.1088/0004-637X/758/1/27}, \href
  {https://ui.adsabs.harvard.edu/abs/2012ApJ...758...27L} {758, 27}

\bibitem[\protect\citeauthoryear{{Liang}, {Zhang}, {Virgili}  \& {Dai}}{{Liang}
  et~al.}{2007}]{Liang2007}
{Liang} E.,  {Zhang} B.,  {Virgili} F.,   {Dai} Z.~G.,  2007, \mn@doi [\apj]
  {10.1086/517959}, \href
  {https://ui.adsabs.harvard.edu/abs/2007ApJ...662.1111L} {662, 1111}

\bibitem[\protect\citeauthoryear{{Lipunov} et~al.,}{{Lipunov}
  et~al.}{2010}]{Lipunov2010}
{Lipunov} V.,  et~al., 2010, \mn@doi [Advances in Astronomy]
  {10.1155/2010/349171}, \href
  {https://ui.adsabs.harvard.edu/abs/2010AdAst2010E..30L} {2010, 349171}

\bibitem[\protect\citeauthoryear{{Lipunov} et~al.,}{{Lipunov}
  et~al.}{2022}]{Lipunov2022}
{Lipunov} V.~M.,  et~al., 2022, \mn@doi [Universe] {10.3390/universe8050271},
  \href {https://ui.adsabs.harvard.edu/abs/2022Univ....8..271L} {8, 271}

\bibitem[\protect\citeauthoryear{{Littlejohns} et~al.,}{{Littlejohns}
  et~al.}{2015}]{Littlejohns2015}
{Littlejohns} O.~M.,  et~al., 2015, \mn@doi [\mnras] {10.1093/mnras/stv479},
  \href {https://ui.adsabs.harvard.edu/abs/2015MNRAS.449.2919L} {449, 2919}

\bibitem[\protect\citeauthoryear{{MacFadyen} \& {Woosley}}{{MacFadyen} \&
  {Woosley}}{1999}]{MacFadyen1999}
{MacFadyen} A.~I.,  {Woosley} S.~E.,  1999, \mn@doi [\apj] {10.1086/307790},
  \href {https://ui.adsabs.harvard.edu/abs/1999ApJ...524..262M} {524, 262}

\bibitem[\protect\citeauthoryear{{Magnier} et~al.,}{{Magnier}
  et~al.}{2020}]{Magnier2020}
{Magnier} E.~A.,  et~al., 2020, \mn@doi [\apjs] {10.3847/1538-4365/abb82a},
  \href {https://ui.adsabs.harvard.edu/abs/2020ApJS..251....6M} {251, 6}

\bibitem[\protect\citeauthoryear{{Metzger}, {Giannios}, {Thompson},
  {Bucciantini}  \& {Quataert}}{{Metzger} et~al.}{2011}]{Metzger2011}
{Metzger} B.~D.,  {Giannios} D.,  {Thompson} T.~A.,  {Bucciantini} N.,
  {Quataert} E.,  2011, \mn@doi [\mnras] {10.1111/j.1365-2966.2011.18280.x},
  \href {https://ui.adsabs.harvard.edu/abs/2011MNRAS.413.2031M} {413, 2031}

\bibitem[\protect\citeauthoryear{{Modjaz} et~al.,}{{Modjaz}
  et~al.}{2008}]{Modjaz2008}
{Modjaz} M.,  et~al., 2008, \mn@doi [\aj] {10.1088/0004-6256/135/4/1136}, \href
  {https://ui.adsabs.harvard.edu/abs/2008AJ....135.1136M} {135, 1136}

\bibitem[\protect\citeauthoryear{{Monet} et~al.,}{{Monet}
  et~al.}{2003}]{Monet2003}
{Monet} D.~G.,  et~al., 2003, \mn@doi [\aj] {10.1086/345888}, \href
  {https://ui.adsabs.harvard.edu/abs/2003AJ....125..984M} {125, 984}

\bibitem[\protect\citeauthoryear{{Nousek} et~al.,}{{Nousek}
  et~al.}{2006}]{Nousek2006}
{Nousek} J.~A.,  et~al., 2006, \mn@doi [\apj] {10.1086/500724}, \href
  {https://ui.adsabs.harvard.edu/abs/2006ApJ...642..389N} {642, 389}

\bibitem[\protect\citeauthoryear{{Nouvel de la Fl{\`e}che} et~al.,}{{Nouvel de
  la Fl{\`e}che} et~al.}{2022}]{Nouvel2022}
{Nouvel de la Fl{\`e}che} A.,  et~al., 2022, in {Holland} A.~D.,  {Beletic} J.,
   eds,  Society of Photo-Optical Instrumentation Engineers (SPIE) Conference
  Series Vol. 12191, X-Ray, Optical, and Infrared Detectors for Astronomy X. p.
  121910Q (\mn@eprint {arXiv} {2209.00386}), \mn@doi{10.1117/12.2627826}

\bibitem[\protect\citeauthoryear{{Nouvel de la Fl{\`e}che} et~al.,}{{Nouvel de
  la Fl{\`e}che} et~al.}{2023}]{Nouvel2023}
{Nouvel de la Fl{\`e}che} A.,  et~al., 2023, \mn@doi [arXiv e-prints]
  {10.48550/arXiv.2306.03716}, \href
  {https://ui.adsabs.harvard.edu/abs/2023arXiv230603716N} {p. arXiv:2306.03716}

\bibitem[\protect\citeauthoryear{{O'Connor} et~al.,}{{O'Connor}
  et~al.}{2023}]{OConnor2023}
{O'Connor} B.,  et~al., 2023, \mn@doi [arXiv e-prints]
  {10.48550/arXiv.2302.07906}, \href
  {https://ui.adsabs.harvard.edu/abs/2023arXiv230207906O} {p. arXiv:2302.07906}

\bibitem[\protect\citeauthoryear{{Paczynski}}{{Paczynski}}{1986}]{Paczynski1986}
{Paczynski} B.,  1986, \mn@doi [\apjl] {10.1086/184740}, \href
  {https://ui.adsabs.harvard.edu/abs/1986ApJ...308L..43P} {308, L43}

\bibitem[\protect\citeauthoryear{{Paczynski}}{{Paczynski}}{1991}]{Paczynski1991}
{Paczynski} B.,  1991, \actaa, \href
  {https://ui.adsabs.harvard.edu/abs/1991AcA....41..257P} {41, 257}

\bibitem[\protect\citeauthoryear{{Panter}, {Heavens}  \& {Jimenez}}{{Panter}
  et~al.}{2004}]{Panter2004}
{Panter} B.,  {Heavens} A.~F.,   {Jimenez} R.,  2004, \mn@doi [\mnras]
  {10.1111/j.1365-2966.2004.08355.x}, \href
  {https://ui.adsabs.harvard.edu/abs/2004MNRAS.355..764P} {355, 764}

\bibitem[\protect\citeauthoryear{{Pereyra} et~al.,}{{Pereyra}
  et~al.}{2022}]{Pereyra2022}
{Pereyra} M.,  et~al., 2022, \mn@doi [\mnras] {10.1093/mnras/stac389}, \href
  {https://ui.adsabs.harvard.edu/abs/2022MNRAS.511.6205P} {511, 6205}

\bibitem[\protect\citeauthoryear{{Perley} et~al.,}{{Perley}
  et~al.}{2014}]{Perley2014}
{Perley} D.~A.,  et~al., 2014, \mn@doi [\apj] {10.1088/0004-637X/781/1/37},
  \href {https://ui.adsabs.harvard.edu/abs/2014ApJ...781...37P} {781, 37}

\bibitem[\protect\citeauthoryear{{Petrosian}, {Kitanidis}  \&
  {Kocevski}}{{Petrosian} et~al.}{2015}]{Petrosian2015}
{Petrosian} V.,  {Kitanidis} E.,   {Kocevski} D.,  2015, \mn@doi [\apj]
  {10.1088/0004-637X/806/1/44}, \href
  {https://ui.adsabs.harvard.edu/abs/2015ApJ...806...44P} {806, 44}

\bibitem[\protect\citeauthoryear{{Planck Collaboration} et~al.,}{{Planck
  Collaboration} et~al.}{2014}]{Planck2014}
{Planck Collaboration} et~al., 2014, \mn@doi [\aap]
  {10.1051/0004-6361/201321529}, \href
  {https://ui.adsabs.harvard.edu/abs/2014A&A...571A...1P} {571, A1}

\bibitem[\protect\citeauthoryear{{Porciani} \& {Madau}}{{Porciani} \&
  {Madau}}{2001}]{Porciani2001}
{Porciani} C.,  {Madau} P.,  2001, \mn@doi [\apj] {10.1086/319027}, \href
  {https://ui.adsabs.harvard.edu/abs/2001ApJ...548..522P} {548, 522}

\bibitem[\protect\citeauthoryear{{Rees} \& {Meszaros}}{{Rees} \&
  {Meszaros}}{1994}]{Rees1994}
{Rees} M.~J.,  {Meszaros} P.,  1994, \mn@doi [\apjl] {10.1086/187446}, \href
  {https://ui.adsabs.harvard.edu/abs/1994ApJ...430L..93R} {430, L93}

\bibitem[\protect\citeauthoryear{{Ryan}, {van Eerten}, {Piro}  \&
  {Troja}}{{Ryan} et~al.}{2020}]{Ryan2020}
{Ryan} G.,  {van Eerten} H.,  {Piro} L.,   {Troja} E.,  2020, \mn@doi [\apj]
  {10.3847/1538-4357/ab93cf}, \href
  {https://ui.adsabs.harvard.edu/abs/2020ApJ...896..166R} {896, 166}

\bibitem[\protect\citeauthoryear{{Santana}, {Barniol Duran}  \&
  {Kumar}}{{Santana} et~al.}{2014}]{Santana2014}
{Santana} R.,  {Barniol Duran} R.,   {Kumar} P.,  2014, \mn@doi [\apj]
  {10.1088/0004-637X/785/1/29}, \href
  {https://ui.adsabs.harvard.edu/abs/2014ApJ...785...29S} {785, 29}

\bibitem[\protect\citeauthoryear{{Sari}, {Piran}  \& {Narayan}}{{Sari}
  et~al.}{1998}]{Sari1998}
{Sari} R.,  {Piran} T.,   {Narayan} R.,  1998, \mn@doi [\apjl]
  {10.1086/311269}, \href
  {https://ui.adsabs.harvard.edu/abs/1998ApJ...497L..17S} {497, L17}

\bibitem[\protect\citeauthoryear{{Savaglio}}{{Savaglio}}{2006}]{Savaglio2006}
{Savaglio} S.,  2006, \mn@doi [New Journal of Physics]
  {10.1088/1367-2630/8/9/195}, \href
  {https://ui.adsabs.harvard.edu/abs/2006NJPh....8..195S} {8, 195}

\bibitem[\protect\citeauthoryear{{Schlegel}, {Finkbeiner}  \&
  {Davis}}{{Schlegel} et~al.}{1998}]{Schlegel1998}
{Schlegel} D.~J.,  {Finkbeiner} D.~P.,   {Davis} M.,  1998, \mn@doi [\apj]
  {10.1086/305772}, \href
  {https://ui.adsabs.harvard.edu/abs/1998ApJ...500..525S} {500, 525}

\bibitem[\protect\citeauthoryear{{Skrutskie} et~al.,}{{Skrutskie}
  et~al.}{2006}]{Skrutskie2006}
{Skrutskie} M.~F.,  et~al., 2006, \mn@doi [\aj] {10.1086/498708}, \href
  {https://ui.adsabs.harvard.edu/abs/2006AJ....131.1163S} {131, 1163}

\bibitem[\protect\citeauthoryear{{Smith} et~al.,}{{Smith}
  et~al.}{2021}]{Smith2021}
{Smith} K.~L.,  et~al., 2021, \mn@doi [\apj] {10.3847/1538-4357/abe6a2}, \href
  {https://ui.adsabs.harvard.edu/abs/2021ApJ...911...43S} {911, 43}

\bibitem[\protect\citeauthoryear{{Srinivasaragavan}, {Dainotti}, {Fraija},
  {Hernandez}, {Nagataki}, {Lenart}, {Bowden}  \& {Wagner}}{{Srinivasaragavan}
  et~al.}{2020}]{Srinivasaragavan2020}
{Srinivasaragavan} G.~P.,  {Dainotti} M.~G.,  {Fraija} N.,  {Hernandez} X.,
  {Nagataki} S.,  {Lenart} A.,  {Bowden} L.,   {Wagner} R.,  2020, \mn@doi
  [\apj] {10.3847/1538-4357/abb702}, \href
  {https://ui.adsabs.harvard.edu/abs/2020ApJ...903...18S} {903, 18}

\bibitem[\protect\citeauthoryear{{Steele} et~al.,}{{Steele}
  et~al.}{2017}]{Steele2017}
{Steele} I.~A.,  et~al., 2017, \mn@doi [\apj] {10.3847/1538-4357/aa79a2}, \href
  {https://ui.adsabs.harvard.edu/abs/2017ApJ...843..143S} {843, 143}

\bibitem[\protect\citeauthoryear{{Thompson}}{{Thompson}}{1994}]{Thompson1994}
{Thompson} C.,  1994, \mn@doi [\mnras] {10.1093/mnras/270.3.480}, \href
  {https://ui.adsabs.harvard.edu/abs/1994MNRAS.270..480T} {270, 480}

\bibitem[\protect\citeauthoryear{{Tody}}{{Tody}}{1986}]{Tody1986}
{Tody} D.,  1986, in {Crawford} D.~L.,  ed.,  Society of Photo-Optical
  Instrumentation Engineers (SPIE) Conference Series Vol. 627, Instrumentation
  in astronomy VI. p.~733, \mn@doi{10.1117/12.968154}

\bibitem[\protect\citeauthoryear{{Urrutia}, {De Colle}, {Murguia-Berthier}  \&
  {Ramirez-Ruiz}}{{Urrutia} et~al.}{2021}]{Urrutia2021}
{Urrutia} G.,  {De Colle} F.,  {Murguia-Berthier} A.,   {Ramirez-Ruiz} E.,
  2021, \mn@doi [\mnras] {10.1093/mnras/stab723}, \href
  {https://ui.adsabs.harvard.edu/abs/2021MNRAS.503.4363U} {503, 4363}

\bibitem[\protect\citeauthoryear{{Vestrand} et~al.,}{{Vestrand}
  et~al.}{2002}]{Vestrand2002}
{Vestrand} W.~T.,  et~al., 2002, in {Kibrick} R.~I.,  ed.,  Society of
  Photo-Optical Instrumentation Engineers (SPIE) Conference Series Vol. 4845,
  Advanced Global Communications Technologies for Astronomy II. pp 126--136
  (\mn@eprint {arXiv} {astro-ph/0209300}), \mn@doi{10.1117/12.459515}

\bibitem[\protect\citeauthoryear{{Virgili}, {Zhang}, {Nagamine}  \&
  {Choi}}{{Virgili} et~al.}{2011}]{Virgili2011}
{Virgili} F.~J.,  {Zhang} B.,  {Nagamine} K.,   {Choi} J.-H.,  2011, \mn@doi
  [\mnras] {10.1111/j.1365-2966.2011.19459.x}, \href
  {https://ui.adsabs.harvard.edu/abs/2011MNRAS.417.3025V} {417, 3025}

\bibitem[\protect\citeauthoryear{{Wanderman} \& {Piran}}{{Wanderman} \&
  {Piran}}{2010}]{Wanderman2010}
{Wanderman} D.,  {Piran} T.,  2010, \mn@doi [\mnras]
  {10.1111/j.1365-2966.2010.16787.x}, \href
  {https://ui.adsabs.harvard.edu/abs/2010MNRAS.406.1944W} {406, 1944}

\bibitem[\protect\citeauthoryear{{Wang}, {Liang}, {Li}, {Lu}, {Wei}  \&
  {Zhang}}{{Wang} et~al.}{2013}]{Wang2013}
{Wang} X.-G.,  {Liang} E.-W.,  {Li} L.,  {Lu} R.-J.,  {Wei} J.-Y.,   {Zhang}
  B.,  2013, \mn@doi [\apj] {10.1088/0004-637X/774/2/132}, \href
  {https://ui.adsabs.harvard.edu/abs/2013ApJ...774..132W} {774, 132}

\bibitem[\protect\citeauthoryear{{Watson} et~al.,}{{Watson}
  et~al.}{2012}]{Watson2012}
{Watson} A.~M.,  et~al., 2012, in {Stepp} L.~M.,  {Gilmozzi} R.,   {Hall}
  H.~J.,  eds,  Society of Photo-Optical Instrumentation Engineers (SPIE)
  Conference Series Vol. 8444, Ground-based and Airborne Telescopes IV. p.
  84445L, \mn@doi{10.1117/12.926927}

\bibitem[\protect\citeauthoryear{{Watson} et~al.,}{{Watson}
  et~al.}{2016}]{Watson2016}
{Watson} A.~M.,  et~al., 2016, in {Evans} C.~J.,  {Simard} L.,   {Takami} H.,
  eds,  Society of Photo-Optical Instrumentation Engineers (SPIE) Conference
  Series Vol. 9908, Ground-based and Airborne Instrumentation for Astronomy VI.
  p. 99085O (\mn@eprint {arXiv} {1606.00690}), \mn@doi{10.1117/12.2233000}

\bibitem[\protect\citeauthoryear{{Woosley}}{{Woosley}}{1993}]{Woosley1993}
{Woosley} S.~E.,  1993, \mn@doi [\apj] {10.1086/172359}, \href
  {https://ui.adsabs.harvard.edu/abs/1993ApJ...405..273W} {405, 273}

\bibitem[\protect\citeauthoryear{{Wright}}{{Wright}}{2006}]{Wright2006}
{Wright} E.~L.,  2006, \mn@doi [\pasp] {10.1086/510102}, \href
  {https://ui.adsabs.harvard.edu/abs/2006PASP..118.1711W} {118, 1711}

\bibitem[\protect\citeauthoryear{{Yu}, {Wang}, {Dai}  \& {Cheng}}{{Yu}
  et~al.}{2015}]{Yu2015}
{Yu} H.,  {Wang} F.~Y.,  {Dai} Z.~G.,   {Cheng} K.~S.,  2015, \mn@doi [\apjs]
  {10.1088/0067-0049/218/1/13}, \href
  {https://ui.adsabs.harvard.edu/abs/2015ApJS..218...13Y} {218, 13}

\bibitem[\protect\citeauthoryear{{Zhang}}{{Zhang}}{2018}]{Zhang2018}
{Zhang} B.,  2018, {The Physics of Gamma-Ray Bursts}.
Cambridge University Press, \mn@doi{10.1017/9781139226530}

\bibitem[\protect\citeauthoryear{{Zhang} \& {Yan}}{{Zhang} \&
  {Yan}}{2011}]{Zhang2011}
{Zhang} B.,  {Yan} H.,  2011, \mn@doi [\apj] {10.1088/0004-637X/726/2/90},
  \href {https://ui.adsabs.harvard.edu/abs/2011ApJ...726...90Z} {726, 90}

\bibitem[\protect\citeauthoryear{{Zhang}, {Fan}, {Dyks}, {Kobayashi},
  {M{\'e}sz{\'a}ros}, {Burrows}, {Nousek}  \& {Gehrels}}{{Zhang}
  et~al.}{2006}]{Zhang2006}
{Zhang} B.,  {Fan} Y.~Z.,  {Dyks} J.,  {Kobayashi} S.,  {M{\'e}sz{\'a}ros} P.,
  {Burrows} D.~N.,  {Nousek} J.~A.,   {Gehrels} N.,  2006, \mn@doi [\apj]
  {10.1086/500723}, \href
  {https://ui.adsabs.harvard.edu/abs/2006ApJ...642..354Z} {642, 354}

\makeatother
\end{thebibliography}

\begin{comment}
\begin{longtable}{lrrrrrr}
\caption{GRBs properties} \label{tab:properties}
\hline
GRB& Trigger& Extinction& Redshift&Duration&&\\
& (BAT)& E(B-V)&z&\multicolumn{3}{l}{$T_{90}$ [s]} \\ 
\hline
\endfirsthead
\multicolumn{7}{c}%
{{\bfseries \tablename\ \thetable{} -- continued from previous page}} \\
\hline
GRB& Trigger& Extinction& Redshift&Duration&&\\
& (BAT)& E(B-V)&z&$T_{90}$ [s]&& \\ 
\hline 
\endhead
\hline 
\endfoot
\hline \hline
\endlastfoot
\input{tablanuevainfo.tex}
\hline
\end{longtable}
\end{comment}

\begin{table*}
	\centering
	\caption{COATLI Observations of GRB 180706A} 
	\label{tab:grb180706A}
\begin{tabular}{ccr}
\hline
$T$ ($s$) & Exposure ($s$)& $w$ (AB)\\
\hline
615& 5& 17.07$\pm$ 0.14\\
624& 5& 17.17$\pm$ 0.13\\
634& 5& 17.11$\pm$ 0.13\\
\end{tabular}
\end{table*}

\begin{table*}
	\centering
	\caption{COATLI Observations of GRB 180812A} 
	\label{tab:grb180812A}
\begin{tabular}{ccr}
\hline
$T$ ($s$) & Exposure ($s$)& $w$ (AB)\\
\hline
46	&	5	&	16.82	$\pm$	0.14	\\
55	&	5	&	16.79	$\pm$	0.13	\\
65	&	5	&	16.70	$\pm$	0.12	\\
74	&	5	&	16.83	$\pm$	0.14	\\
83	&	5	&	16.85	$\pm$	0.14	\\
92	&	5	&	16.74	$\pm$	0.13	\\
101	&	5	&	16.52	$\pm$	0.11	\\
110	&	5	&	16.61	$\pm$	0.12	\\
119	&	5	&	16.76	$\pm$	0.13	\\
128	&	5	&	16.61	$\pm$	0.11	\\
141	&	5	&	16.76	$\pm$	0.13	\\
150	&	5	&	16.59	$\pm$	0.11	\\
159	&	5	&	16.73	$\pm$	0.12	\\
168	&	5	&	16.59	$\pm$	0.11	\\
177	&	5	&	16.72	$\pm$	0.12	\\
186	&	5	&	16.79	$\pm$	0.13	\\
195	&	5	&	16.73	$\pm$	0.12	\\
204	&	5	&	16.75	$\pm$	0.12	\\
215	&	5	&	16.58	$\pm$	0.11	\\
224	&	5	&	17.03	$\pm$	0.16	\\
237	&	5	&	16.76	$\pm$	0.13	\\
246	&	5	&	17.01	$\pm$	0.16	\\
255	&	5	&	17.19	$\pm$	0.19	\\
264	&	5	&	16.95	$\pm$	0.15	\\
273	&	5	&	17.05	$\pm$	0.16	\\
282	&	5	&	17.21	$\pm$	0.19	\\
291	&	5	&	17.22	$\pm$	0.19	\\
300	&	5	&	17.74	$\pm$	0.31	\\
309	&	5	&	17.16	$\pm$	0.18	\\
318	&	5	&	17.25	$\pm$	0.20	\\
331	&	5	&	17.07	$\pm$	0.17	\\
340	&	5	&	17.30	$\pm$	0.21	\\
349	&	5	&	17.23	$\pm$	0.19	\\
358	&	5	&	17.36	$\pm$	0.22	\\
367	&	5	&	17.53	$\pm$	0.25	\\
376	&	5	&	17.32	$\pm$	0.21	\\
385	&	5	&	17.40	$\pm$	0.23	\\
394	&	5	&	17.45	$\pm$	0.23	\\
403	&	5	&	17.19	$\pm$	0.19	\\
412	&	5	&	17.33	$\pm$	0.21	\\
426	&	5	&	17.25	$\pm$	0.20	\\
\end{tabular}
\end{table*}

\newpage

\tablefirsthead{\multicolumn{5}{c}{{\bfseries  Table 3.} Properties of GRBs used in this work\label{tab:properties}
}\\
\midrule
GRB&Trigger&E(B-V)&z&$T_{90}$ (s)\\ \midrule}
\tablehead{%
\multicolumn{4}{c}%
{{\bfseries  Table 3. (Continued)}} \\
\midrule
GRB&Trigger&E(B-V)&z&$T_{90}$ (s)\\ \midrule}
\tabletail{
\midrule}
\begin{supertabular}{lrrrr}
GRB 050525A	&	130088	&	$...$	&	...	&	8.83	$\pm$	0.07\\
GRB 060124	&	178750	&	0.13	&	...	&	13.42	$\pm$	1.29	\\
GRB 060209	&	180931	&	0.93	&	...		& ...		\\
GRB 060512	&	209755	&	0.02	&	2.10	&	8.40	$\pm$	1.73	\\
GRB 060515	&	210084	&	0.03	&	...	&	52.37	$\pm$	5.48	\\
GRB 060814	&	224552	&	0.04	&	1.92	&	145.07	$\pm$	4.82	\\
GRB 060825	&	226382	&	0.57	&	...	&	7.98	$\pm$	0.66	\\
GRB 060904A	&	227996	&	0.02	&	2.55	&	80.06	$\pm$	2.22	\\
GRB 060904B	&	228006	&	0.17	&	0.70	&	189.98	$\pm$	21.16	\\
GRB 060919	&	230115	&	0.07	&	...	&	9.00	$\pm$	1.67	\\
GRB 061019	&	234516	&	1.14	&	...	&	180.38	$\pm$	6.36	\\
GRB 061027	&	235645	&	0.03	&	...	&	105.81	$\pm$	36.94	\\
GRB 061028	&	235715	&	0.16	&	...	&	105.81	$\pm$	36.94	\\
GRB 061109	&	237821	&	0.19	&	...			& ...	\\
GRB 061110B	&	238174	&	0.09	&	3.43	&	135.25	$\pm$	17.59	\\
GRB 061217	&	251634	&	0.04	&	0.83	&	0.22	$\pm$	0.04	\\
GRB 061218	&	251863	&	0.12 &	...	& ...			\\
GRB 061222B	&	252593	&	0.41	&	3.35	&	37.25	$\pm$	6.06	\\
GRB 070103	&	254532	&	0.07	&	2.62	&	18.41	$\pm$	1.21	\\
GRB 070227	&	262347	&	0.22	&	...	&	3.20	$\pm$	1.60	\\
GRB 070330	&	273180	&	0.06	&	...	&	6.64	$\pm$	0.67	\\
GRB 070411	&	275087	&	0.10	&	2.95	&	115.69	$\pm$	16.87	\\
GRB 070412	&	275119	&	0.02	&	...	&	33.88	$\pm$	5.08	\\
GRB 070420	&	276321	&	0.52	&	3.01	&	77.02	$\pm$	5.47	\\
GRB 070508	&	278854	&	0.03	&	0.82	&	20.9	$\pm$	0.73	\\
GRB 070521	&	279935	&	0.03 &	2.09	&	38.63	$\pm$	2.38	\\
GRB 070610	&	281993	&	0.81	&	...			& ...	\\
GRB 070621	&	282808	&	0.05	&	...	&	33.26	$\pm$	2.51	\\
GRB 070628	&	283320	&	0.77	&	...	&	39.08	$\pm$	4.07	\\
GRB 070913	&	290843	&	0.13	&	...	&	2.68	$\pm$	0.56	\\
GRB 070920A	&	291614	&	0.10	&	...	&	55.76	$\pm$	5.29	\\
GRB 070920B	&	291728	&	0.01	&	...	&	22.26	$\pm$	2.79	\\
GRB 071010A	&	293707	&	0.42	&	0.98	&	6.32	$\pm$	1.87	\\
GRB 071101	&	295779	&	1.29	&	...	&	4.82	$\pm$	1.97	\\
GRB 071112C	&	296504	&	1.29	&	0.82	&	44.8	$\pm$	1.60	\\
GRB 080129	&	301981	&	1.02	&	4.35	&	50.18	$\pm$	9.17	\\
GRB 080207	&	302728	&	0.02	&	2.08	&	292.46	$\pm$	7.99	\\
GRB 080210	&	302888	&	0.08	&	2.64	&	42.26	$\pm$	11.37	\\
GRB 080315	&	306323	&	0.13	&	...			&	... \\
GRB 080330	&	308041	&	0.02	&	1.51	&	60.36	$\pm$	36.4	\\
GRB 080413A	&	309096	&	0.16	&	2.43	&	46.36	$\pm$	0.48	\\
GRB 080430	&	310613	&	0.01	&	0.77	&	13.87	$\pm$	1.90	\\
GRB 080516	&	311762	&	0.35	&	3.20	&	5.76	$\pm$	0.22	\\
GRB 080603B	&	313087	&	0.01	&	2.69	&	59.12	$\pm$	1.63	\\
GRB 080727C	&	318170	&	0.07	&	...	&	77.36	$\pm$	7.04	\\
GRB 080903	&	323542	&	0.22	&	...	&	66.34	$\pm$	10.42	\\
GRB  081003B	&	325080	&	0.64	&	...	&	30.00	$\pm$		...\\
GRB 081028A	&	332851	&	0.03	&	3.04	&	284.42	$\pm$	30.55	\\
GRB 081109A	&	334112	&	0.02	&	0.98	&	221.49	$\pm$	58.97	\\
GRB 081126	&	335647	&	0.68	&	...	&	57.65	$\pm$	5.85	\\
GRB 081228	&	338338	&	0.16	&	3.44	&	3.00	$\pm$	1.41	\\
GRB 090102	&	338895	&	0.03	&	1.55	&	28.32	$\pm$	2.35	\\
GRB 090307A	&	345551	&	0.53	&	...	&	22.00	$\pm$	6.40	\\
GRB 090313	&	346386	&	0.03	&	3.37	&	83.04	$\pm$	19.48	\\
GRB 090519	&	352648	&	0.04	&	3.85	&	58.04	$\pm$	8.18	\\
GRB 091127	&	377179	&	0.04	&	0.49	&	6.96	$\pm$	0.15	\\
GRB 100316A	&	416076	&	0.03	&	3.15	&	6.75	$\pm$	0.95	\\
GRB 100816A	&	431764	&	0.09	&	0.80	&	2.88	$\pm$	0.63	\\
GRB 100915A	&	434178	&	0.51	&	...	&	199.20	$\pm$	26.19	\\
GRB 100917A	&	434360	&	0.09	&	...	&	63.82	$\pm$	21.56	\\
GRB 110128A	&	443861	&	0.01	&	2.34	&	14.15	$\pm$	2.49	\\
GRB 110205A	&	444643	&	0.01	&	2.22	&	249.42	$\pm$	15.03	\\
GRB 110207A	&	444912	&	0.03	&	...	&	82.58	$\pm$	10.83	\\
GRB 110422A	&	451901	&	0.03	&	1.77	&	25.78	$\pm$	0.60	\\
GRB 110520A	&	453747	&	0.03	&	...	&	20.86	$\pm$	5.37	\\
GRB 111209A	&	509336	&	0.02	&	0.68	&	810.97	$\pm$	52.11	\\
GRB 111225A	&	510341	&	0.26	&	0.30	&	105.73	$\pm$	26.18	\\
GRB 111228A	&	510649	&	0.03	&	0.72	&	101.24	$\pm$	5.45	\\
GRB 120106A	&	511235	&	1.19	&	...	&	63.50	$\pm$	4.62	\\
GRB 120119A	&	512035	&	0.11	&	1.73	&	68.00	$\pm$	7.07	\\
GRB 120324A	&	518507	&	1.12	&	1.10	&	118.42	$\pm$	10.01	\\
GRB 120326A	&	518626	&	0.05	&	1.80	&	69.48	$\pm$	8.18	\\
GRB 120327A	&	518731	&	0.34	&	2.81	&	63.53	$\pm$	7.03	\\
GRB 120328A	&	518792	&	0.73	&	...	&	29.93	$\pm$	6.67	\\
GRB 120403B	&	519256	&	0.13	&	...	&	7.28	$\pm$	1.92	\\
GRB 120514A	&	522197	&	1.63	&	...	&	164.34	$\pm$	5.84	\\
GRB 120521A	&	522578	&	0.43	&	...	&	0.51	$\pm$	0.14	\\
GRB 120521B	&	522586	&	0.45	&	...	&	146.83	$\pm$	27.94	\\
GRB 120521C	&	522656	&	0.01	&	5.93	&	27.07	$\pm$	4.34	\\
GRB 120805A	&	530031	&	0.03	&	3.10	&	48.00	$\pm$	22.63	\\
GRB 120807A	&	530267	&	1.41	&	...	&	19.88	$\pm$	5.76	\\
GRB 120811C	&	530689	&	0.03	&	2.67	&	24.34	$\pm$	3.06	\\
GRB 120815A	&	531003	&	0.10	&	2.36	&	7.23	$\pm$	2.52	\\
GRB 120816A	&	531223	&	0.96	&	...	&	4.43	$\pm$	0.98	\\
GRB 120909A	&	533060	&	0.09	&	3.93	&	220.6	$\pm$	304.97	\\
GRB 120913B	&	533613	&	0.06	&	...	&	122.59	$\pm$	4.46	\\
GRB 120922A	&	534394	&	0.15	&	3.10	&	168.22	$\pm$	29.24	\\
GRB 120923A	&	534402	&	0.16	&	7.84	&	26.08	$\pm$	6.82	\\
GRB 121024A	&	536580	&	0.1	&	2.30	&	67.97	$\pm$	31.72	\\
GRB 121108A	&	537921	&	0.37	&	...	&	89.55	$\pm$	47.54	\\
GRB 121202A	&	540255	&	0.05	&	...	&	19.54	$\pm$	1.49	\\
GRB 130131B	&	547420	&	0.03	&	2.54	&	4.30	$\pm$	0.26	\\
GRB 130215A	&	548760	&	0.10	&	0.60	&	66.22	$\pm$	10.66	\\
GRB 130216A	&	548927	&	0.49	&	...	&	6.52	$\pm$	1.09	\\
GRB 130327A	&	552063	&	0.13	&	...	&	6.37	$\pm$	1.41	\\
GRB 130418A	&	553847	&	0.03	&	1.22	&	274.92	$\pm$	39.32	\\
GRB 130420A	&	553977	&	0.01	&	1.30	&	121.14	$\pm$	11.75	\\
GRB 130427A	&	554620	&	0.24	&	0.34	&	244.33	$\pm$	4.73	\\
GRB 130505A	&	555163	&	0.04	&	2.27	&	89.34	$\pm$	111.15	\\
GRB 130514A	&	555821	&	0.23	&	3.60	&	214.19	$\pm$	17.32	\\
GRB 130521A	&	556344	&	0.52	&	...	&	10.96	$\pm$	2.50	\\
GRB 130603B	&	557310	&	0.02	&	0.36	&	0.18	$\pm$	0.02	\\
GRB 130605A	&	557508	&	0.36	&	...	&	10.18	$\pm$	3.20\\
GRB 130606A	&	557589	&	0.02	&	5.91	&	276.66	$\pm$	19.63	\\
GRB 130610A	&	557845	&	0.02	&	2.09	&	47.72	$\pm$	10.74	\\
GRB 130612A	&	557976	&	0.08	&	2.01	&	4.00	$\pm$	1.41	\\
GRB 130615A	&	558271	&	0.12	&	...	&	332.56	$\pm$	28.55	\\
GRB 130722A	&	563213	&	0.63	&	...	&	98.37	$\pm$	12.86	\\
GRB 130907A	&	569992	&	0.01	&	1.24	&	364.37	$\pm$	5.45	\\
GRB 130925A	&	571830	&	0.02	&	0.35	&	160.30	$\pm$	3.39	\\
GRB 130929A	&	572308	&	2.58	&	...	&	12.24	$\pm$	4.06	\\
GRB 131105A	&	576738	&	0.03	&	1.69	&	112.21	$\pm$	4.10	\\
GRB 131117A	&	577968	&	0.02	&	4.04	&	10.88	$\pm$	2.81	\\
GRB 140114A	&	583861	&	0.02	&	3.00	&	139.95	$\pm$	15.88	\\
GRB 140215A	&	586680	&	0.09	&	...	&	25.63	$\pm$	5.64	\\
GRB 140302A	&	589685	&	0.52	&	...	&	98.72	$\pm$	11.40	\\
GRB 140304A	&	590206	&	0.08	&	5.28	&	14.78	$\pm$	1.40	\\
GRB 140311A	&	591390	&	0.03	&	4.95	&	70.48	$\pm$	7.59	\\
GRB 140320A	&	592544	&	0.51	&	...	&	0.51	$\pm$	0.23	\\
GRB 140331A	&	594081	&	0.05	&	4.65	&	209.66	$\pm$	32.75	\\
GRB 140413A	&	595616	&	0.02	&	...	&	132.98	$\pm$	14.37	\\
GRB 140419A	&	596426	&	0.03	&	3.96	&	80.08	$\pm$	3.78	\\
GRB 140423A	&	596901	&	0.01	&	3.26	&	134.14	$\pm$	23.10	\\
GRB 140516A	&	599188	&	0.02	&	...	&	0.26	$\pm$	0.14	\\
GRB 140518A	&	599287	&	0.02	&	4.71	&	60.52	$\pm$	2.48	\\
GRB 140610A	&	601259	&	2.56	&	...	&	93.25	$\pm$	3.35	\\
GRB 140622A	&	602278	&	0.07	&	0.96	&	0.13	$\pm$	0.04	\\
GRB 140626A	&	602604	&	0.15	&	...	&	16.16	$\pm$	1.23	\\
GRB 140703A	&	603243	&	0.10	&	3.14	&	68.64	$\pm$	66.46	\\
GRB 140709A	&	603810	&	0.40	&	...	&	105.19	$\pm$	7.39	\\
GRB 140710A	&	603954	&	0.06	&	0.56	&	3.00	$\pm$	2.24	\\
GRB 141004A	&	614390	&	0.32	&	0.57	&	3.92	$\pm$	1.11	\\
GRB 141015A	&	615399	&	0.51	&	...	&	11.00	$\pm$	4.12	\\
GRB 141022A	&	616061	&	0.10	&	...	&	8.72	$\pm$	1.56	\\
GRB 141031A	&	617110	&	0.17	&	...	&	464.00	$\pm$	472.27	\\
GRB 141109A	&	618024	&	0.04	&	2.99	&	200.19	$\pm$	48.15	\\
GRB 141121A	&	619182	&	0.05	&	1.47	&	481.00	$\pm$	38.06	\\
GRB 141221A	&	622006	&	0.03	&	1.45	&	36.82	$\pm$	4.10	\\
GRB 141225A	&	622476	&	0.02	&	0.91	&	86.15	$\pm$	45.22	\\
GRB 150203A	&	629578	&	0.90	&	...	&	24.44	$\pm$	2.77	\\
GRB 150213B	&	631051	&	0.02	&	...	&	209.00	$\pm$	74.24	\\
GRB 150323A	&	635887	&	0.03	&	0.59	&	149.73	$\pm$	9.12	\\
GRB 150323C	&	636005	&	0.01	&	...	&	159.66	$\pm$	44.62	\\
GRB 150716A	&	649157	&	1.43	&	...	&	42.57	$\pm$	5.59	\\
GRB 150727A	&	650530	&	...	&	0.31	&	87.96	$\pm$	10.99	\\
GRB 150728A	&	650617	&	0.20	&	...	&	0.83	$\pm$	0.23	\\
GRB 150811A	&	651882	&	0.17	&	...	&	31.52	$\pm$	7.32	\\
GRB 150817A	&	652334	&	0.48	&	...	&	38.00	$\pm$	2.13	\\
GRB 151215A	&	667392	&	0.39	&	2.59	&	17.85	$\pm$	1.01	\\
GRB 160117B	&	670800	&	0.07	&	0.87	&	11.54	$\pm$	2.61	\\
GRB 160121A	&	671231	&	0.53	&	1.96	&	10.50	$\pm$	2.40	\\
GRB 160123A	&	671447	&	0.10	&	...	&	3.95	$\pm$	0.62	\\
GRB 160127A	&	671828	&	0.06	&	...	&	6.16	$\pm$	0.94	\\
GRB 160131A	&	672236	&	0.11	&	0.97	&	327.75	$\pm$	71.03	\\
GRB 160203A	&	672525	&	0.07	&	3.52	&	17.44	$\pm$	2.29	\\
GRB 160220A	&	674670	&	0.13	&	...	&	8.26	$\pm$	1.17	\\
GRB 160225A	&	675998	&	0.01	&	...	&	157.46	$\pm$	70.49	\\
GRB 160228A	&	676595	&	0.06	&	1.64	&	98.91	$\pm$	23.89	\\
GRB 160313A	&	678929	&	0.01	&	...	&	41.54	$\pm$	6.18	\\
GRB 160314A	&	679120	&	0.08	&	0.73	&	8.73	$\pm$	1.52	\\
GRB 160321A	&	680017	&	0.75	&	...	&	33.48	$\pm$	4.83	\\
GRB 160327A	&	680655	&	0.01	&	4.99	&	33.74	$\pm$	9.37	\\
GRB 160410A	&	682269	&	0.02	&	1.72	&	96.00	$\pm$	50.6	\\
GRB 160501A	&	684679	&	0.18	&	...	&	118.00	$\pm$	19.00	\\
GRB 160504A	&	685124	&	0.01	&	...	&	53.96	$\pm$	5.68	\\
GRB 160506A	&	685245	&	0.26	&	...	&	260.53	$\pm$	78.32	\\
GRB 160630A	&	702252	&	0.48	&	...	&	29.55	$\pm$	5.61	\\
GRB 160705B	&	703176	&	0.01	&	...	&	54.58	$\pm$	12.57	\\
GRB 160714A	&	704310	&	0.02	&	...	&	0.35	$\pm$	0.11	\\
GRB 160821A	&	709351	&	0.03	&	...	&	112.64	$\pm$	1.10	\\
GRB 161004A	&	715084	&	0.25	&	...	&	1.32	$\pm$	0.30	\\
GRB 161014A	&	717500	&	0.08	&	2.82	&	23.00	$\pm$	4.47	\\
GRB 161022A	&	718655	&	0.04	&	...	&	7.25	$\pm$	1.07	\\
GRB 161105A	&	720697	&	0.33	&	...	&	177.70	$\pm$	25.07	\\
GRB 161108A	&	721234	&	0.02	&	0.50	&	115.84	$\pm$	11.66	\\
GRB 161117A	&	722604	&	0.06	&	1.55	&	125.70	$\pm$	1.09	\\
GRB 161214B	&	726885	&	0.08	&	$<$1.5	&	24.90	$\pm$	3.18	\\
GRB 161217A	&	727167	&	0.01	&	...	&	12.04	$\pm$	2.45	\\
GRB 161218A	&	727288	&	...		&	... &7.20	$\pm$	0.55	\\
GRB 170112A	&	732188	&	0.02	&	$<$0.2\footnote{\cite{Anderson2018}}	&	0.06	$\pm$	0.02	\\
GRB 170202A	&	736407	&	0.02	&	3.64	&	37.76	$\pm$	11.72	\\
GRB 170317A	&	742866	&	0.19	&	...	&	11.95	$\pm$	4.87	\\
GRB 170318B	&	743086	&	4.51	&	...	&	1.07	$\pm$	0.22	\\
GRB 170405A	&	745797	&	0.09	&	3.51	&	165.31	$\pm$	32.89	\\
GRB  170409A	&	748858	&	...		&... &	68	$\pm$	...	\\
GRB 170428A	&	750298	&	0.06	&	0.45	&	0.20	$\pm$	0.07	\\
GRB 170519A	&	753445	&	0.03	&	0.82	&	220.25	$\pm$	143.84	\\
GRB 170604A	&	755867	&	0.04	&	1.33	&	26.53	$\pm$	2.84	\\
GRB 170705A	&	760064	&	0.02	&	2.01	&	223.20	$\pm$	15.55	\\
GRB 170823A	&	769177	&	...		&	...& 69.41	$\pm$	12.03	\\
GRB 170921A	&	773509	&	0.05	&	...	&	28.80	$\pm$	5.93	\\
GRB 171011A	&	778154	&	...		&	...&2.28	$\pm$	0.23	\\
GRB 171013A	&	778435	&	0.64	&	...	&	18.90	$\pm$	1.50	\\
GRB 171020A	&	780845	&	0.19	&	1.87	&	41.86	$\pm$	9.19	\\
GRB 171205A	&	794972	&	0.05	&	0.04	&	190.47	$\pm$	33.88	\\
GRB 180111A	&	804692	&	0.01	&	...	&	50.89	$\pm$	1.88	\\
GRB 180205A	&	808625	&	0.03	&	1.41	&	15.54	$\pm$	3.15	\\
GRB 180224A	&	811561	&	0.01	&	...	&	10.92	$\pm$	3.57	\\
GRB 180314A	&	814129	&	0.08	&	1.44	&	50.46	$\pm$	21.84	\\
GRB 180316A	&	814677	&	0.28	&	...	&	87.00	$\pm$	20.59	\\
GRB 180325A	&	817564	&	0.02	&	2.25	&	92.83	$\pm$	1.52	\\
GRB 180329B	&	819490	&	0.04	&	2.00	&	213.54	$\pm$	49.42	\\
GRB 180418A	&	826428	&	0.02	&	...	&	4.41	$\pm$	2.49	\\
GRB 180510B	&	831816	&	0.26	&	1.30	&	134.02	$\pm$	52.92	\\
GRB 180620A	&	843122	&	0.13	&	$<$1.2	&	22.95	$\pm$	4.99	\\
GRB 180624A	&	844192	&	0.05	&	2.85	&	466.65	$\pm$	35.76	\\
GRB 180626A	&	844615	&	0.04	&	...	&	30.10	$\pm$	1.41	\\
GRB 180705A	&	846299	&	0.06	&	...	&	106.78	$\pm$	16.2	\\
GRB 180706A	&	846395	&	0.02	&	...	&	42.44	$\pm$	7.16	\\
GRB 180809B	&	852553	&	0.16	&	...	&	233.24	$\pm$	1.12	\\
GRB 180812A	&	852903	&	0.04	&	...	&	16.53	$\pm$	0.97	\\
GRB 180904A	&	859282	&	0.04	&	...	&	4.63	$\pm$	1.03	\\
GRB 190427A	&	900730	&	0.06	&	...	&	0.19	$\pm$	0.20	\\
GRB 191011A	&	928924	&	0.02	&	1.72	&	7.35	$\pm$	0.94	\\
GRB 191016A	&	929744	&	0.09	&	3.29\footnote{\cite{Smith2021}}	&	126.68	$\pm$	7.73\\
GRB 191031A	&	932435 & ... &  ... & 19.10 $\pm$ 6.30 \\
GRB 191031D	&	932608 & 0.08 &  ... & 0.29 $\pm$ 0.05 \\
GRB 200122A	&	950330	&	0.03	&	...	&	190.60	$\pm$	4.51	\\
GRB 200125A	&	952164	&	0.08	&	...	&	350.34	$\pm$	32.97	\\
GRB 201021C	&	1001130	&	0.02	&	1.07	&	24.38	$\pm$	6.76	\\
GRB 210222B	&	1034325	&	0.10	&	2.20	&	12.11	$\pm$	0.91	\\
GRB 210504A	&	1046782	&	...	&	2.08&	142.98	$\pm$	9.33	\\
GRB 210610A	&	1054627	&	0.03    &	3.54	&	13.62	$\pm$	3.15	\\
GRB 210610B	&	1054681	&	0.04	&	1.13	&	69.38	$\pm$	2.53	\\
GRB 210722A	&	1061223	&	0.03	&	1.14	&	50.20	$\pm$	10.54	\\
GRB 210822A	&	1069788	&	0.14	&	1.74	&	185.68	$\pm$	46.63	\\
GRB 220118A &   1093742 & 0.03 & ... & 10.61 $\pm$ 1.78\\
 \\
\end{supertabular}

\label{lastpage}
\end{document}